\documentclass[12pt]{article}
\usepackage{amsmath,amssymb,exscale,cancel}
\usepackage{slashed}
\allowdisplaybreaks
\usepackage{graphicx}

\usepackage{graphicx}
\usepackage{epsfig}
\usepackage{multicol}
\usepackage{color}
\usepackage{mathrsfs}
\usepackage{blindtext}
 \usepackage{fancyhdr}
\usepackage{hyperref}
\usepackage{cite}
\usepackage{mathtools}

\usepackage{floatrow}

\usepackage{graphicx}
\usepackage{sidecap}

\usepackage[latin1]{inputenc} 

\usepackage{multicol}  
 
 \usepackage{titlesec}
 
 \usepackage{rotating,slashed,xcolor,amsfonts,expdlist,charter}

\numberwithin{equation}{section}

\usepackage{xcolor}
\usepackage{sectsty}

\usepackage{mdframed}
\usepackage{titletoc}

\numberwithin{equation}{section}

\usepackage{xcolor}
\usepackage{sectsty}

\usepackage{mdframed}
\usepackage{titletoc}

\definecolor{secnum}{RGB}{13,151,225}
\definecolor{ptcbackground}{RGB}{212,237,252}
\definecolor{ptctitle}{RGB}{0,177,235}

\titlecontents{lsection}
  [5.8em]{\sffamily}
  {\color{secnum}\contentslabel{2.3em}\normalcolor}{}
  {\titlerule*[1000pc]{.}\contentspage\\\hspace*{-5.8em}\vspace*{5pt}%
    \color{white}\rule{\dimexpr\textwidth-15.5pt\relax}{1pt}}

\usepackage{hyperref}
\hypersetup{colorlinks,bookmarksopen,bookmarksnumbered,citecolor=rossos,
linkcolor=redy,pdfstartview=FitH,urlcolor=green-go}
\usepackage{slashed}

\definecolor{blus}{cmyk}{1,0.9,0,0.1}
\definecolor{verdes}{cmyk}{0.99,0,0.59,0.65}
\definecolor{rossos}{cmyk}{0,1,1,0.55}
\definecolor{redy}{cmyk}{0,1,1,0.7}
\definecolor{greeny}{cmyk}{0.99,0,0.59,0.98}
\definecolor{green-go}{cmyk}{0.79,0,0.59,0.5}

\usepackage{titlesec}

\newcommand{\beq}{\begin{equation}}
\newcommand{\eeq}{\end{equation}}

\def\hhref#1{\href{http://arxiv.org/abs/#1}{arXiv:#1}} 

\newcommand{\tmtextbf}[1]{{\bfseries{#1}}}
\newcommand{\tmtextrm}[1]{{\rmfamily{#1}}}
\newcommand{\bp}{\bar M_{Pl}}

\def\be{\begin{equation}}
\def\ee{\end{equation}}
\def\ba{\begin{array} }

\def\bac{\begin{array} {c}}
\def\bacc{\begin{array} {cc}}
\def\baccc{\begin{array} {ccc}}
\def\bacccc{\begin{array} {cccc}}
\def\ea{\end{array}}
\def\bea{\begin{eqnarray}}
\def\eea{\end{eqnarray}}

\definecolor{red}{rgb}{1,0,0}

\def\psl{\hbox{\hbox{${p}$}}\kern-1.9mm{\hbox{${/}$}}}
\def\dsl{\hbox{\hbox{${\partial}$}}\kern-2.2mm{\hbox{${/}$}}}
\def\Dsl{\hbox{\hbox{${D}$}}\kern-2.6mm{\hbox{${/}$}}}

\newcommand{\gappeq}{{\rlap{{\raise}.5ex\text{\ensuremath{>}}}{{\lower}.5ex\text{\ensuremath{\sim}}}}}
\newcommand{\lappeq}{{\rlap{{\raise}.5ex\text{\ensuremath{<}}}{{\lower}.5ex\text{\ensuremath{\sim}}}}}
\newcommand{\I}{\tmtextrm{1{\kern}-.24em l}}

\begin{document}
\topmargin -1.0cm
\oddsidemargin 0.9cm
\evensidemargin -0.5cm

{\vspace{-1cm}}
\begin{center}

\vspace{-1cm}

 {\tmtextbf{ 
 \hspace{-1.2cm}   
{\LARGE  \color{rossos}On the Validity of the \\ Effective Theory  of (Multi-)Field Inflation}
 }} {\vspace{.5cm}}\\

\vspace{1.3cm}

{\large{\bf  Andrea Ambrosi de Magistris$^{a}$} and {\bf  Alberto Salvio$^{b,c}$ }}

{\em  
\vspace{.4cm}

$^{a}$ Physics Department, Sapienza University of Roma, Piazzale Aldo Moro 5, 00185, Roma, Italy

\vspace{0.6cm}

$^{b}$ Physics Department, University of Rome Tor Vergata, \\ 
via della Ricerca Scientifica, I-00133 Rome, Italy\\

\vspace{0.6cm}

$^{c}$ I. N. F. N. -  Rome Tor Vergata,\\
via della Ricerca Scientifica, I-00133 Rome, Italy\\

  \vspace{0.5cm}

}
\vspace{1.5cm}
\end{center}

\noindent ---------------------------------------------------------------------------------------------------------------------------------
\begin{center}
{\bf \large Abstract}
\end{center}
\noindent   Motivated by trans-Planckian issues in inflation, we determine the Hilbert space and amplitudes of quantum perturbations in the general low-energy effective theory of (multi-)field inflation without relying on the sub-horizon limit. The scalar sector is the most intricate, featuring field mixings and second-class constraints, which we handle using Dirac brackets. These results enable us to estimate the magnitude of higher-derivative corrections. In the specific case of slow-roll inflation, such estimate can be expressed in terms of the first slow-roll parameter $\epsilon$ for a given cutoff $\Lambda$. We apply our results to several inflationary models with finite $\Lambda$: Higgs inflation, the Starobinsky model, natural inflation and hilltop inflation.

\vspace{0.7cm}

\noindent---------------------------------------------------------------------------------------------------------------------------------

\newpage

\tableofcontents

\noindent --------------------------------------------------------------------------------------------------------------------------------

\vspace{0.2cm}

\section{Introduction}\label{intro}

Inflation is the current paradigm for the very early universe as it leads to predictions in agreement with cosmic-microwave-background data for several models~\cite{Ade:2015lrj,ACT:2025fju,ACT:2025tim}. An interesting aspect is that inflation provides us with a window on very energetic phenomena and could give us information on the ultraviolet completion of the Standard Model and Einstein gravity. 

Nevertheless, inflationary models are typically specific implementations of the general logic of effective field theories (EFTs)~\cite{Donoghue:1994dn,Donoghue:2017ovt,Weinberg:2008hq,Khosravi:2012qg}, which feature a finite energy cutoff  $\Lambda$  above which physical consistency is lost. It is therefore important to clearly understand the limits of applicability of EFTs in the inflationary context. On the one hand, this understanding is necessary to consistently extract predictions from specific EFT models of inflation. On the other hand, it may provide clues as to where observational signatures of UV completions could arise.

\begin{figure}[t!]
  \centering
  \includegraphics[width=0.48\textwidth]{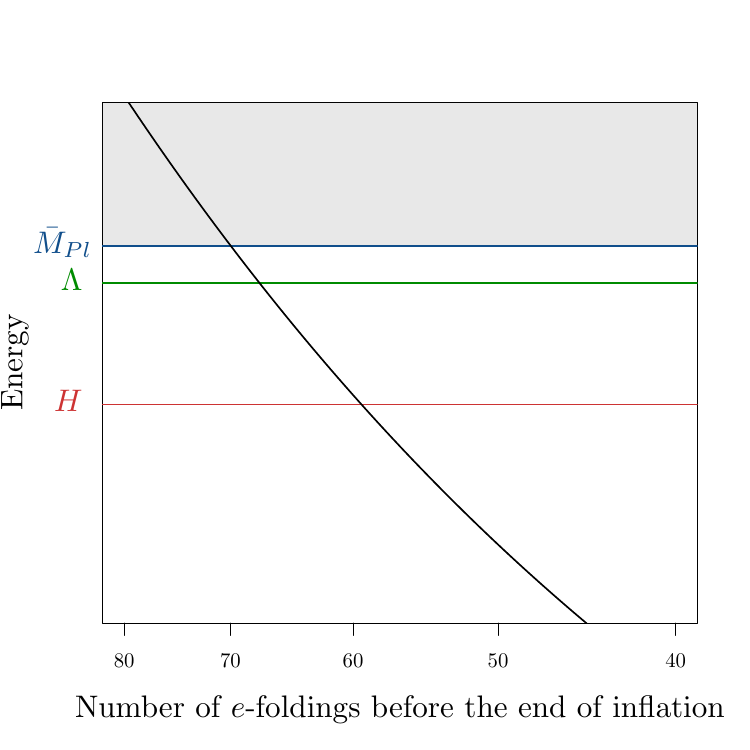}
  \caption{\em Qualitative plot of the evolution of $q/a$ during inflation. The number of e-folds reported is provided for illustrative purposes only.}
  \label{Fig:qaPlot}
\end{figure}

Several works have pointed out a number of trans-Planckian issues in inflationary EFTs (see e.g.~\cite{Martin:2000xs,Brandenberger:2000wr,Brandenberger:2022pqo}). Some of these problems arise from the fact that, when the exponential inflationary expansion is extrapolated backward in time, the physical momentum $q/a$ of a perturbation eventually becomes comparable to $\Lambda$, whose maximal value is the reduced Planck mass $\bp$. Even if the inflationary Hubble rate $H$ is much smaller than $\Lambda$, the EFT breaks down in the sub-horizon limit ($q/a\gg H$) when $q/a\gtrsim\Lambda$. This situation is illustrated schematically in Fig.~\ref{Fig:qaPlot}. But the fact that simple extrapolations might cause
problems cannot be used as an argument to dismiss the entire EFT, which should be tested within its domain
of validity only~\cite{Chowdhury:2019otk,Burgess:2020nec}.

However, the sub-horizon limit is widely employed in the literature within the canonical quantization procedure of (multi-)field inflationary theories  (see e.g.~\cite{weinberg2008,LangloisRenauxPetel2008}). If taking $q/a\gg H$ is physically necessary to determine the Hilbert space and the amplitude of quantum fluctuations, the accuracy of the EFT is reduced: calling $n$ the minimal number of extra derivatives that are neglected in a given use of the EFT, the relative uncertainty increases from $(H/\Lambda)^n$ up to $(q/(a\Lambda))^n$ computed at the earliest time that one must consider.

To clarify this situation we study here the quantization of inflationary perturbations without relying on the sub-horizon limit. We perform a general analysis going beyond the slow-roll approximation and considering the general relativistic low-energy EFT of (multi-)field inflation. Allowing for an arbitrary number of scalars is phenomenologically important because, except in the case of inflation within the pure Standard Model~\cite{Bezrukov:2007ep}, multiple scalars can generically take part in the dynamics. Existing treatments commonly employ the sub-horizon limit. The scalar sector features complicated mixings and constraints, which however become trivial when $q/a$ is sufficiently larger than $H$. Moreover, the same limit is used in the literature to uniquely determine the Hilbert space and amplitude of all quantum perturbations. Understanding to what extent this limit is physically necessary to recover the common inflationary predictions is the main goal of this paper.

Here we also estimate the size of the EFT relative uncertainty $(q/(a\Lambda))^n$ in the ordinary scenario of new inflation providing famous and observationally successful  models as key examples.

The paper is organized as follows. In the next section we define the relativistic low-energy EFT of (multi-)field inflation and establish our conventions. Since vector perturbations are trivial, in Secs.~\ref{Tensor perturbations} and~\ref{Scalar perturbations} we focus on tensor and scalar perturbations, respectively. The intricate mixings and constraints appear in the scalar sector. In Sec.~\ref{Applications} we provide some examples of UV completions and the estimate of the size of the EFT relative uncertainty in the general new inflation scenario, providing the examples of Higgs inflation, the Starobinsky model, natural inflation and hilltop inflation. Finally, in Sec.~\ref{conclusions} we discuss related issues, the choice of the inflationary quantum state and unitarity, summarize the main findings of the paper and present some outlook. The reader may also read Sec.~\ref{conclusions} first to have a quick overview of the paper.

\section{The low energy effective field theory}\label{eft}

We consider the low-energy relativistic inflationary EFT that includes in the set of degrees of freedom the massless spin-2 graviton and $N$ spin-0 particles with fields $\phi\equiv\{\phi^1,\ldots,\phi^N\}$, which are conventionally taken to be real. The leading action in the low-energy expansion reads\footnote{We use the signature $(+1,-1,-1,-1)$ for the space-time metric, $g_{\mu\nu}$, define  $g$ to be the determinant of the space-time metric and $$R_{\mu\nu\,\,\, \sigma}^{\quad \rho} \equiv \partial_{\mu} \Gamma_{\nu \, \sigma}^{\,\rho}- \partial_{\nu} \Gamma_{\mu \, \sigma}^{\,\rho} +  \Gamma_{\mu \, \tau}^{\,\rho}\Gamma_{\nu \,\sigma }^{\,\tau}- \Gamma_{\nu \, \tau}^{\,\rho}\Gamma_{\mu \,\sigma }^{\,\tau}, \quad \Gamma_{\mu \, \sigma}^{\,\rho} \equiv \frac12 g^{\rho \tau}\left(\partial_{\mu} g_{\sigma\tau}+\partial_{\sigma} g_{\mu\tau}-\partial_{\tau} g_{\mu\sigma}\right), \quad R_{\mu\nu} \equiv R_{\rho\mu\,\,\, \nu}^{\quad \rho},\quad R\equiv g^{\mu\nu}R_{\mu\nu}.$$
Also,  we use natural units where $c=\hbar=1$.}
\begin{equation}
    S_{ES} = \int d^4 x\ \sqrt{-g}\ \bigg[\frac{K_{ij}(\phi)}{2}\partial_{\mu}\phi^i \partial^{\mu} \phi^j - U(\phi)-\frac{\bp^2}{2}R\bigg],
    \label{eq:SehMultifield}
\end{equation}
where $K_{ij}$ represents the metric on the space of scalar fields; it can be taken to be symmetric without loss of generality and is assumed here to be positive definite to avoid ghosts at low energies. The case of a flat scalar-field-space is recovered by simply choosing $K_{ij} = \delta_{ij}$. In our notation, we distinguish the indices $i,j, ...$ in their upper and lower positions, and the tensor $K_{ij}$ and its inverse $K^{ij}$ will be used, respectively, to lower and raise these indices.  We will not make any specific assumption about the potential density $U$ in this section and in  Secs.~\ref{Tensor perturbations} and~\ref{Scalar perturbations}, but we will provide some examples in Sec.~\ref{Applications}. 

Action~(\ref{eq:SehMultifield}) is the most general two-derivative action describing the graviton and the $N$ spin-0 particles. Indeed, any additional non-minimal coupling of the form $\sqrt{-g} f(\phi)R$ can be reabsorbed in a redefinition of $K_{ij}$ and $U$ through a Weyl transformation of the metric. The original field variables, where non-minimal couplings are present, form the Jordan frame, while the new field variables, where the action is expressed like in~(\ref{eq:SehMultifield}), is known as the Einstein frame. Higher-derivative terms are assumed to be suppressed by inverse powers of $\Lambda$, the EFT cutoff. The next-to-leading terms in the low-energy expansion feature four derivatives and include for example $R^2$, $R_{\mu\nu}R^{\mu\nu}$, $K_{ij}\phi^i\Box^2\phi^j$, where $\Box\equiv \nabla_\mu \nabla^\mu$ is the covariant d'Alembertian, etc..

Various methods to study multi-field inflationary theories with the general action in~\ref{eq:SehMultifield} have been provided~\cite{Starobinsky:1985ibc,Sasaki:1995aw,Nakamura:1996da,Gordon:2000hv,Sasaki:2008uc,Chiba:2008rp,White:2012ya,Kaiser:2012ak,Kaiser:2013sna,White:2013ufa}. To describe inflation in this EFT we expand the metric around a Friedmann-Lema\^itre-Robertson-Walker 
(FLRW) background\footnote{We neglect  the spatial curvature in the FLRW metric as the energy density is expected to be dominated by the inflatons during inflation. Also, the curvature contribution to the background field equations decreases extremely rapidly as the universe expands exponentially.}:
\vspace{2mm}
\begin{equation}
    ds^2 = a(\eta)^2 \left\{ (1 + 2\Phi(\eta, \vec{x}))\, d\eta^2 
- 2V_i(\eta, \vec{x})\, d\eta\, dx^i 
- \left[(1 - 2\Psi(\eta, \vec{x}))\delta_{ij} + h_{ij}(\eta, \vec{x})\right] dx^i dx^j \right\},
\label{eq:ds2}
\end{equation}
with $\eta$ being the conformal time defined in terms of the scale factor $a$ and the cosmic time $t$ by
\begin{equation}
    \eta = \int_{t_0}^{t} \frac{dt'}{a(t')},
    \label{eq:conformalt ime}
\end{equation}
where $t_0$ is  some reference cosmic time.
We have adopted the conformal Newtonian gauge with metric scalar ($\Phi$, $\Psi$), vector ($V_i$) and tensor ($h_{ij}$) perturbations.
Vector and tensor perturbations satisfy
\begin{equation}
    \partial_i V_i = 0,\hspace{1cm}h_{ij} = h_{ji},\hspace{1cm}h_{ii} = 0,\hspace{1cm}\partial_ih_{ij} = 0.
    \label{eq:metric conditions}
\end{equation}
Correspondingly, the inflaton fields can be expressed as the sum of a space-independent background and small perturbations:
\begin{equation}
    \phi^i(\vec{x},\eta)\to \phi^i(\eta) + \varphi^i (\vec{x},\eta).
    \label{eq:inflaton expansion}
\end{equation}

At the background level the Einstein equations  are   \bea {\cal H}^2&=&\frac{K_{ij} \phi'^i \phi'^j/2+a^2U}{3 \bar M_{\rm Pl}^2} ,\label{EOM1c} \\ 
{\cal H}^2-{\cal H}'&=& \frac{K_{ij} \phi'^i \phi'^j}{ 2\bar M_{\rm Pl}^2}, \label{eq:phi'2MF}\eea
 where ${\cal H} \equiv a'/a$ (related to the Hubble rate $H\equiv \dot a/a$ by ${\cal H} = aH$) and a prime denotes a derivative with respect to $\eta$, while a dot is a derivative with respect to $t$. The equations for the scalar fields are instead
 \be \phi''^i +\gamma^i_{jk} \phi'^j \phi'^k +2{\cal H} \phi'^i+a^2U^{,i}=0. \label{eq:phi''MF}\ee
 Here  for a generic function $F$ of the scalar fields, we defined $F_{,i}\equiv \partial F/\partial \phi^i$ and $F^{,i}\equiv K^{ij}F_{,j}$, also $\gamma^i_{jk}$ is the affine connection in the scalar field space
 \be \gamma^i_{jk}\equiv \frac{K^{il}}{2}\left(K_{lj,k}+K_{lk,j}-K_{jk,l}\right). \ee
 
 Notice that in the case of pure de Sitter space-time ($a(\eta)=-1/(H\eta)$, $\phi'^i=0,$ $U_{,i} =0$) Eq. (\ref{EOM1c}) tells us 
  $ {\cal H}^2 =  a^2 U/(3\bp^2)$ and Eq.~(\ref{eq:phi'2MF}) implies 
  ${\cal H}' = {\cal H}^2$.
  
  Vector perturbations satisfy the linearized equations $\vec{\nabla}^2V_i\equiv\partial_j\partial_j V_i=0$ and are therefore not generated. In the rest of the paper we can thus focus on tensor and scalar perturbations, which we analyze in turn.

\section{Tensor perturbations}\label{Tensor perturbations}

At linearized level tensor and scalar perturbations are not coupled. The procedure for the quantization of tensor perturbations is much simpler than that for scalar perturbations, because tensor perturbations (unlike scalars) do not feature mixings and can be reduced to an unconstrained system. We therefore start from the tensor sector. 

Eq.~\eqref{eq:SehMultifield} leads to the following action $S_{EH}^{(T)}$ quadratic in tensor perturbations 
\begin{equation}
S_{EH}^{(T)} = \frac{\bp^2}{8} \int d^4x\, a^2 \left( h_{ij}' h_{ij}' + h_{ij} \vec\nabla^2 h_{ij} \right).
\label{eq:ActionTensorPerturbations}
\end{equation}
We can decompose the fields $ h_{ij}$ by performing a spatial Fourier transform and using the polarization tensors $e^\lambda_{ij} (\hat q)$, where the label $\lambda$ indicate here the possible values of the helicities, $\pm 2$: 
\bea h_{ij}(\vec{x},\eta) =  \int \frac{d^3q}{(2\pi)^3} \, e^{i\vec{q}\cdot \vec{x}} \sum_{\lambda= \pm 2}  h_\lambda(\vec{q},\eta) e^\lambda_{ij} (\hat q),  \label{ExpMom}  \eea
with $\hat q\equiv \vec q/q$ and $q\equiv |\vec q|$. Then the tensor field equations read
\begin{equation}
    h''_\lambda + 2\mathcal{H}h_\lambda' +q^2 h_\lambda = 0.
    \label{eq:heq}
\end{equation}

We recall that for $\hat q$ along the third axis the polarization tensors that satisfy the last three conditions in (\ref{eq:metric conditions}) are  given by 
\be e^{+2}_{11} = -e^{+2}_{22} = 1/2, \quad e^{+2}_{12} = e^{+2}_{21} = i/2, \quad e^{+2}_{3i} =e^{+2}_{i3} = 0, \quad  e^{-2}_{ij} =  (e^{+2}_{ij})^* \label{PolT}\ee 
and for a generic momentum direction $\hat q$ we can obtain $e^\lambda_{ij} (\hat q)$ by applying to (\ref{PolT}) a rotation that connects the third axis with  $\hat q$. The polarization tensors then obey 
\be e^\lambda_{ij} (\hat q) e^{*\lambda'}_{ij} (\hat q) = \delta^{\lambda\lambda'}  \label{eq:enormalization}\ee 
and 
\begin{equation}
    e^{\lambda}_{ij}(-\hat{q}) = e^{-\lambda}_{ij}(\hat{q}).
    \label{eq:e-qProperty}
\end{equation}
Note that the Hermiticity of $h_{ij}(\vec x,\eta)$ implies $h^\dagger_\lambda(\vec q,\eta)= h_\lambda(-\vec q,\eta)$.

We also introduce
\begin{equation}
\Lambda_{ij,kl}(\hat{q})\equiv \sum_{\lambda} e^{\lambda}_{ij}(\hat{q})\  e^{*\lambda}_{kl}(\hat{q}).
\label{eq:LambdaDefinition}
\end{equation}
One can easily check that $e^{-\lambda}_{ij} =  e^{*\lambda}_{ij}$ and~(\ref{eq:e-qProperty}) imply
\begin{equation}
\Lambda_{ij,kl}(\hat{q}) = \Lambda_{kl,ij}(\hat{q}), \qquad \Lambda_{ij,kl}(-\hat{q}) = \Lambda_{ij,kl}(\hat{q})
\label{eq:LambdaProperties}
\end{equation}
and \eqref{eq:enormalization} leads to
\begin{equation}
 e^{*\lambda}_{ij}(\hat{q})\ \Lambda_{ij,kl}(\hat{q})\  e^{\lambda'}_{kl}(\hat{q}) = \delta_{\lambda \lambda'}.
\label{eq:LambdaNormalization}
\end{equation}

Eq.~\eqref{eq:heq} admits a pair of complex-conjugate solutions, $h_q$ and $h_q^*$, which depend on $\vec q$ only though $q$. Therefore, the expansion  in~(\ref{ExpMom}) of the Hermitian field $h_{ij}$ can be expressed as follows
\begin{equation}
    h_{ij}(\vec{x},\eta) = \sum_{\lambda = \pm2} \int \frac{d^3q}{(2\pi)^3}  \Bigr[e_{ij}^{\lambda}(\hat{q}) \ h_q(\eta)\  e^{i\vec{q}\cdot\vec{x}}\ b_\lambda(\vec{q}) +e_{ij}^{*\lambda}(\hat{q})\ h_q^*(\eta)\ e^{-i\vec{q}\cdot\vec{x}}\  b^\dagger_\lambda(\vec{q})\Bigl]
    \label{eq:hexp}
\end{equation}
in terms of the tensor creation and annihilation operators, $b^\dagger_\lambda(\vec{q})$ and $b_\lambda(\vec{q})$, respectively.

Now, in order uniquely identify the commutation relations between $b^\dagger_\lambda(\vec{q})$ and $b_\lambda(\vec{q})$ without performing the sub-horizon limit, we invert this expansion to obtain $b_\lambda(\vec{q})$ expressed as a functional of $h_{ij}$ and $ h_{ij}'$. To this end we multiply both sides of Eq.~\eqref{eq:hexp} by $e^{-i\vec{k}\cdot\vec{x}}$ and integrate over $d^3x$ to get: 
\begin{align}
    \int d^3x\ e^{-i\vec{k}\cdot\vec{x}}\ h_{ij}(\vec{x},\eta) = \sum_{\lambda = \pm2}  \Bigr[e_{ij}^{\lambda}(\hat{k}) \  h_k(\eta)\ b_\lambda(\vec{k}) +e_{ij}^{*\lambda}(-\hat{k})\ h_k^*(\eta)  b_\lambda^\dagger(-\vec{k})\Bigl].\nonumber
\end{align}
Note that, according to~\eqref{eq:e-qProperty} and $e^{-\lambda}_{ij} =  e^{*\lambda}_{ij}$, we have $\ e_{ij}^{*\lambda}(-\hat{k}) = e_{ij}^{*\, -\lambda}(\hat{k}) = e_{ij}^{\lambda}(\hat{k})$. This allows the previous equation to be simplified as follows:
\begin{align}   
     \int d^3x\ e^{-i\vec{k}\cdot\vec{x}}\ h_{ij}(\vec{x},\eta) =\sum_{\lambda = \pm2}  e_{ij}^{\lambda}(\hat{k})\Bigr[h_k(\eta)\ b_\lambda(\vec{k}) + h_k^*(\eta)\  b_\lambda^\dagger(-\vec{k})\Bigl].
    \end{align}
Contracting both sides with $e_{ij}^{*\lambda'}(\hat{q})$ and using~\eqref{eq:enormalization}, we obtain
\begin{equation}
    \int d^3x\ e^{-i\vec{q}\cdot\vec{x}}\ e_{ij}^{*\lambda}(\hat{q})\  h_{ij}(\vec{x},\eta) =h_q(\eta)\ b_\lambda(\vec{q}) + h_q^*(\eta)\  b_\lambda^\dagger(-\vec{q})
\end{equation}
and, taking a derivative with respect to $\eta$,
\begin{equation}
    \int d^3x\ e^{-i\vec{q}\cdot\vec{x}}\ e_{ij}^{*\lambda}(\hat{q})\  h'_{ij}(\vec{x},\eta) =h'_q(\eta)\ b_\lambda(\vec{q}) + h_q'^*(\eta)\  b_\lambda^\dagger(-\vec{q}).
\end{equation}
Finally, a linear combination of the last two equations allows to express $b_\lambda(\vec{q})$ as functionals of the fields:
\begin{align}
        \boxed{b_\lambda(\vec{q})=\frac{e_{ij}^{*\lambda}(\hat{q})}{h_q \overset{\leftrightarrow}{\partial_\eta}h_q^*(\eta)}\int d^3x\ e^{-i\vec{q}\cdot\vec{x}}\ h_{ij}(\vec{x},\eta)\ \overset{\leftrightarrow}{\partial_\eta}h_q^*,}       \label{eq:alphahexp}      
\end{align}
where, for any pair $\{f_1, f_2\}$ of functions of $\eta$, we defined $f_1\overset{\leftrightarrow}{\partial_\eta} f_2(\eta)\equiv f_1'(\eta)f_2(\eta)-f_1(\eta)f_2'(\eta)$. Note that the expression in~\eqref{eq:alphahexp} was obtained without taking the sub-horizon limit and in fact holds for {\it any} $\eta$. Since  $b_\lambda(\vec{q})$ is time independent, this  implies that the right-hand side of~\eqref{eq:alphahexp} is also time independent.

Having obtained $b_\lambda(\vec{q})$ as a functional of $h_{ij}$ and $ h_{ij}'$, we can now uniquely identify the commutation relations between $b^\dagger_\lambda(\vec{q})$ and $b_\lambda(\vec{q})$ by imposing the canonical commutation relations, but without performing the sub-horizon limit.  

The canonical quantization cannot be easily applied to the tensor perturbations $h_{ij}(\vec x,\eta)$ since not every component is independent. In fact, the last three conditions in $\eqref{eq:metric conditions}$ significantly reduce the independent components of $h_{ij}$. Note, however, that the decomposition in Eq.~\eqref{ExpMom} already isolates the two physical degrees of freedom. For this reason, it is easier to apply the canonical quantization procedure to the two fields $h_\lambda(\vec{q},\eta)$ with $\lambda = \pm2$.

By using Eq.~\eqref{ExpMom} in~\eqref{eq:ActionTensorPerturbations} we find
\begin{equation}
     S_{EH}^{(T)} = \int d\eta d^3q\, \tilde{\mathscr{L}}_T,\end{equation}
     where 
     \be \tilde{\mathscr{L}}_T \equiv \frac{\bp^2}{8}\sum_{\lambda=\pm2} \frac{a^2(\eta)}{(2\pi)^3}\ \left[ h'_\lambda(\vec{q},\eta)\ h'^\dagger_\lambda(\vec{q},\eta) -q^2\   h_\lambda(\vec{q},\eta)\ h_\lambda^\dagger(\vec{q},\eta) \right] \label{MomLag}
\end{equation}
is the momentum-space Lagrangian density, whose momentum integral gives the Lagrangian $L_T \equiv\int d^3q\, \tilde{\mathscr{L}}_T$.
Then, we can proceed by finding the conjugate momentum from $L_T$ and imposing the canonical commutation rule. Note that, unlike in the scalar sector that we will analyze in Sec.~\ref{Scalar perturbations}, there are no additional constraints on the $h_\lambda$ besides $h^\dagger_\lambda(\vec q,\eta)= h_\lambda(-\vec q,\eta)$. Hence, unlike in the scalar sector, Poisson brackets can immediately be promoted to commutators, which imply: 
\begin{align}
    &\Pi_\lambda(\vec{q},\eta) = \frac{\delta L_T}{\delta h'_\lambda(\vec{q},\eta)} = \frac{a^2(\eta)\bp^2}{4(2\pi)^3}\ h_\lambda'^\dagger(\vec{q},\eta),\label{Pih}\\
    &\Bigl[h_\lambda(\vec{q},\eta), \Pi_{\lambda'}(\vec{k},\eta)\Bigr] = i    \ \delta_{\lambda\lambda'}\ \delta^{(3)}(\vec{q}-\vec{k}),\\
    &\Bigl[h_\lambda(\vec{q},\eta),\ h'^\dagger_{\lambda'}(\vec{k},\eta)\Bigr] = \frac{4 i(2\pi)^3}{a^2(\eta)\bp^2}\ \delta_{\lambda\lambda'}\ \delta^{(3)}(\vec{q}-\vec{k})
    \label{eq:hcommutationrule}
\end{align}
and 
\be \Bigl[h_\lambda(\vec{q},\eta), h_{\lambda'}(\vec{k},\eta)\Bigr] = \Bigl[h'_\lambda(\vec{q},\eta), h'_{\lambda'}(\vec{k},\eta)\Bigr] = 0. \label{hhhhp}\ee

These canonical commutators allow us to determine  the commutators between the Hermitian fields $h_{ij}$ and $h_{kl}'$ by using  Eq.~\eqref{ExpMom} and  its time derivative,
\begin{align}
    \left[h_{ij}(\vec{x},\eta), h'_{kl}(\vec{y},\eta)\right] & = \int \frac{d^3 q}{(2\pi^3)}\frac{d^3 k}{(2\pi)^3}e^{i\vec{q}\cdot\vec{x}-i\vec{k}\cdot\vec{y}}\sum_{\lambda,\lambda'=\pm2} e_{ij}^{\lambda}(\hat{q})  e_{kl}^{*\lambda'}(\hat{k}) \Bigl[h_\lambda(\vec{q},\eta),\ h'^\dagger_{\lambda'}(\vec{k},\eta)\Bigr]\nonumber\\
    & =\frac{4\ i}{a^2(\eta)\bp^2}\ \int \frac{d^3 q}{(2\pi)^3}e^{i\vec{q}\cdot(\vec{x}-\vec{y})}\Lambda_{ij,kl}(\hat{q}), \nonumber
\end{align}
which can then be employed to compute the commutators between creation and annihilation operators through Eq.~\eqref{eq:alphahexp}:
\begin{align*}
    &\left[b_\lambda(\vec{q}),b^\dagger_{\lambda'}(\vec{k})\right] = \frac{e_{ij}^{*\lambda}(\hat{q})}{h_q\overset{\leftrightarrow}{\partial_\eta}h_q^*(\eta)}\ \frac{e_{kl}^{\lambda'}(\hat{k})}{h_k^* \overset{\leftrightarrow}{\partial_\eta}h_k(\eta)}\int d^3x\ d^3y\ e^{-i\vec{q}\cdot\vec{x}+i\vec{k}\cdot\vec{y}}\nonumber\\ 
&\times\left(-h_q^* h'_{k}\left[h'_{ij}(\vec{x},\eta), h_{kl}(\vec{y},\eta)\right]-h'^{*}_q h_{k}\left[h_{ij}(\vec{x},\eta),\ h'_{kl}(\vec{y},\eta)\right]\  \right) \nonumber \\
& =\frac{4\ i}{a^2(\eta)\bp^2} \frac{e_{ij}^{*\lambda}(\hat{q})}{h_q \overset{\leftrightarrow}{\partial_\eta}h_q^*(\eta)}\ \frac{e_{kl}^{\lambda'}(\hat{k})}{h_k^* \overset{\leftrightarrow}{\partial_\eta}h_k(\eta)}\int d^3 x\ d^3 y\ \frac{d^3 p}{(2\pi)^3}\ e^{-i\vec{q}\cdot\vec{x}+i\vec{k}\cdot\vec{y}}\\
&\times \left(h_q^* h'_{k}\Lambda_{kl,ij}(\vec{p})-h'^{*}_q h_{k}\Lambda_{ij,kl}(\vec{p})\  \right)e^{i\vec{p}\cdot(\vec{x}-\vec{y})}\nonumber,
\end{align*}
where we used $\left[h_{ij}(\vec{x},\eta), h_{kl}(\vec{y},\eta)\right]=0$ and $ \left[h'_{ij}(\vec{x},\eta), h'_{kl}(\vec{y},\eta)\right]$, which are simple consequences of~\eqref{hhhhp}.
We can now perform the integrations over $d^3x\ d^3y$, which yield two delta functions, integrate over $d^3 p$ and use the properties of $\Lambda_{ij,kl}$ in Eq. \eqref{eq:LambdaProperties}, to find
\begin{align*}
&\left[b_\lambda(\vec{q}),b_{\lambda'}^\dagger(\vec{k})\right] =\frac{4 i(2\pi)^3}{a^2(\eta)\bp^2} \frac{e_{ij}^{*\lambda}(\hat{q})}{h_q\overset{\leftrightarrow}{\partial_\eta}h_q^*(\eta)}\frac{e_{kl}^{\lambda'}(\hat{q})}{h_q^*\overset{\leftrightarrow}{\partial_\eta}h_q(\eta)} \left(h_q^* h'_q- h'^{*}_q h_q\right)\delta^{(3)}(\vec{q}-\vec{k})\ \Lambda_{ij,kl}(\vec{q}).
\end{align*}
Finally, the property in~\eqref{eq:LambdaNormalization} leads to
\begin{equation}
\left[b_\lambda(\vec{q}),b^\dagger_{\lambda'}(\vec{k})\right] =  \frac{4 (2\pi)^3}{a^2(\eta)\bp^2} \ \frac{i}{h_q^*\overset{\leftrightarrow}{\partial_\eta}h_q(\eta)}\  \delta^{(3)}(\vec{q}-\vec{k})\ \delta_{\lambda \lambda'}.
\label{eq:bbdagcommutator}
\end{equation}
Also, this procedure can be easily  extended to find
    \be  \Big[b_\lambda(\vec{q}),b_{\lambda'}(\vec{k})\Big]=0. \label{Commbb}\ee
Since $b_\lambda(\vec{q})$ is time independent, consistency requires that the right-hand side of Eq.~\eqref{eq:bbdagcommutator} is also time independent\footnote{This can be explicitly checked using Eq.~\eqref{eq:heq}:
\begin{equation*}
    \frac{d}{d\eta}\Bigg[a^2\ h_q^*\overset{\leftrightarrow}{\partial_\eta}h_q\Bigg] = a^2\Bigg[2\mathcal{H}(h_q'^*h_q - h_q^*h_q') + \ (h_q''^*h_q - h_q^*h_q'')\Bigg] = 0.
\end{equation*}
}. Then, we choose a normalization for $h_q$ such that
\begin{equation}
  \frac{4i (2\pi)^3}{a^2\bp^2 \ h_q^*\overset{\leftrightarrow}{\partial_\eta}h_q} = 1 
\label{eq:hnormalization}
\end{equation}
and Eq.~\eqref{eq:bbdagcommutator} implies the usual commutation rules for the tensor creation and annihilation operators:
\begin{equation}
    \Big[b_\lambda(\vec{q}),b_{\lambda'}^\dagger(\vec{k})\Big] =  \delta_{\lambda\lambda'}\ \delta^{(3)}(\vec{q}-\vec{k}). \label{usualCommT}
\end{equation}

Therefore, we have explicitly shown that the Hilbert space 
and the amplitude of quantum fluctuations are determined without relying on the sub-horizon limit. They are determined, respectively, by the commutation rules in~\eqref{usualCommT}-\eqref{Commbb} and the normalization conditions in~\eqref{eq:hnormalization}.

\section{Scalar perturbations}\label{Scalar perturbations}

Let us now move to the most complex sector, that of scalars. As well known, in the conformal Newtonian gauge the linearized field equations for the metric and the scalars lead to $\Phi=\Psi$ and the following set of equations
\begin{subequations}\label{sys:MultiScalPertSys}
\begin{align}
& D^2_\eta\varphi^i  + 2\mathcal{H} D_\eta \varphi^i + a^2 K^{il}U_{,l,j} \varphi^j-\mathcal{F}^{i}_{\ lmn}\phi'^{l}\phi'^{m}\varphi^n -\vec\nabla^2 \varphi^i
=  4 \Psi' \phi'^{i}-2 a^2 K^{i j}U_{,j} \Psi,  \label{eq:MultiScalPertSys1} \\
& \Psi' + \mathcal{H} \Psi
= \frac{1}{2\bp^2} K_{ij} \phi'^{i} \varphi^j,  \label{eq:MultiScalPertSys2} \\ 
&  \Big(\mathcal{H}' - \mathcal{H}^2-\vec\nabla^2\Big)\Psi = \frac{1}{2\bp^2}\ K_{ij} \left( \varphi^iD_\eta\phi'^{j}-\phi'^{i}D_\eta \varphi^j -\mathcal{H}\phi'^{i}\varphi^j \right),\label{eq:MultiScalPertSys3}
\end{align}
\end{subequations}
where $\mathcal{F}^i_{\ lmn}$ is the Riemann tensor in the space of scalars,
\begin{equation}
    \mathcal{F}^i_{\ lmn} = \gamma^i_{ln,m}-\gamma^i_{lm,n} + \gamma^i_{mj}\gamma^j_{ln} - \gamma^i_{nj}\gamma^j_{lm},  
\end{equation}
and $D_\eta$ is the covariant derivative in the same space: its defining effect on a generic vector $v^i$ is
\begin{equation}
    D_\eta v^i = v'^{i} + \gamma^i_{mn}\phi'^{m} v^n.
\end{equation}

On the other hand, inserting the expression of the fields in the conformal Newtonian gauge, Eqs.~(\ref{eq:ds2}) and~\eqref{eq:metric conditions}-\eqref{eq:inflaton expansion}, in the Einstein-scalar action in~(\ref{eq:SehMultifield}) leads to the following scalar action
 \begin{equation}
\begin{aligned}
S^{(S)}_{ES}
&= \int d^4x\, \frac{a^2}{2}
\Bigg\{
\bp^2
\Big[
-6{\Psi'}^2
-12\mathcal{H}\Phi\Psi'
+4\Psi\vec\nabla^2\Phi
-2\Psi\vec\nabla^2\Psi
-2\left(\mathcal{H}' + 2\mathcal{H}^2\right)\Phi^2
\Big] \\
&\quad
+ K_{ij}\left(
\varphi'^{i}\varphi'^{j} + \varphi^i\vec\nabla^2\varphi^j
\right)
+ 2K_{ij,l}\phi'^{i}\varphi'^{j}\varphi^l
- (\Phi + 3\Psi)
\left(
2K_{ij}\phi'^{i}\varphi'^{j} + K_{ij,l}\phi'^{i}\phi'^{j}\varphi^l
\right) \\
&\quad
- a^2 U_{,i,j}\varphi^i\varphi^j
- 2a^2(\Phi - 3\Psi)U_{,i}\varphi^i
\Bigg\}.
\end{aligned} \label{SSES}
\end{equation}
Here we directly use the constraint $\Phi=\Psi$ imposed by the Einstein equations to obtain
\begin{equation}
\begin{aligned}
S^{(S)}_{ES}
&= \int d^4x\, \frac{a^2}{2}
\Bigg\{
\bp^2
\Big[
-6{\Psi'}^2
-12\mathcal{H}\Psi\Psi'
+2\Psi\vec\nabla^2\Psi
-2\left(\mathcal{H}' + 2\mathcal{H}^2\right)\Psi^2
\Big] \\
&\quad
+ K_{ij}\left(
\varphi'^{i}\varphi'^{j} + \varphi^i\vec\nabla^2\varphi^j
\right)
+ 2K_{ij,l}\phi'^{i}\varphi'^{j}\varphi^l
- 4\Psi
\left(
2K_{ij}\phi'^{i}\varphi'^{j} + K_{ij,l}\phi'^{i}\phi'^{j}\varphi^l
\right) \\
&\quad
- a^2 U_{,i,j}\varphi^i\varphi^j
+ 4a^2 \Psi\ U_{,i}\varphi^i
\Bigg\}. \label{eq:SehMultifieldPhiEqPsi}
\end{aligned}
\end{equation}
This simplified action will be our starting point for quantizing the scalar sector. However, it is important to note that not all equations in~\eqref{eq:MultiScalPertSys1}-\eqref{eq:MultiScalPertSys3} can be reproduced by performing the variations of the action in~(\ref{eq:SehMultifieldPhiEqPsi}) with respect to the dynamical variables, $\Psi$ and $\varphi^i$. In particular, Eq.~\eqref{eq:MultiScalPertSys2} cannot be reproduced in this way. The reason is that the procedures of fixing the gauge and deriving the field equations do not generically commute with each other. In particular Eq.~\eqref{eq:MultiScalPertSys2} comes from terms in the action (before fixing the gauge) that contain the divergence of some vector perturbations, like $\partial_i V_i$,
linearly.  If one first fixes the conformal Newtonian gauge (where~(\ref{eq:metric conditions}) holds) these terms just disappear. Instead, if one correctly first performs the variations of the action (before  fixing the gauge) with respect to all dynamical variables~\cite{Mukhanov:1990me}, including $V_i$ and $h_{ij}$, one finds an additional equation from those terms, precisely Eq.~\eqref{eq:MultiScalPertSys2}. Therefore, if one adopts the action in~(\ref{eq:SehMultifieldPhiEqPsi}),  Eq.~\eqref{eq:MultiScalPertSys2} must be imposed as an additional constraint. We will do so in the following\footnote{It is worth mentioning that, if one avoids imposing $\Phi=\Psi$ and works with the action in~(\ref{SSES}), it is possible to derive $\Phi=\Psi$ and all equations in~\eqref{eq:MultiScalPertSys1}-\eqref{eq:MultiScalPertSys3} by performing variations of~(\ref{SSES}) with respect to $\Psi$, $\varphi^j$ {\it and} $\Phi$~\cite{Deruelle:2010kf}. However, as clear from~(\ref{SSES}) in this case one would obtain a field equation for $\Phi$ that is non-dynamical (without time derivative of  $\Phi$) and the canonical momentum conjugate to $\Phi$ identically zero, which means that a constraint would be present anyway.}.  

The system of linear equations under study, Eqs.~\eqref{eq:MultiScalPertSys1}-\eqref{eq:MultiScalPertSys3}, can be analysed, like for the tensor perturbations, in momentum space replacing $\vec\nabla^2$ with $-q^2$. Then there is one second-order differential equation for each real scalar field $\varphi^i$, and one first-order equation for the real scalar $\Psi$ and one constraint on $\Psi$. So the number of real independent solutions for each $q$ is $2N$. We   denote them by ($\varphi^i_{n},\Psi_{n}$) for $n=1,\ldots,2N$, omitting their dependence on $q$. These solutions are not uniquely determined: we have the freedom to perform a change of basis in the space of solutions of the form:
\begin{align}
\varphi^i_n  \to T_{nm}\ \varphi^i_m, \qquad \Psi_{n}\to T_{nm}\ \Psi_{m},
\end{align}
with $T$ being an arbitrary time-independent invertible $2N\times2N$ matrix, and still obtain a valid set of solutions. The freedom to choose a basis in the $2N$-dimensional solution space will be exploited later.

\subsection{Constrained Hamiltonian analysis}\label{Constrained Hamiltonian analysis}

To canonically quantize this system we now introduce the conjugate momenta of the fields $\Psi$ and $\varphi^i$  from the Lagrangian density $\mathscr{L}^{(S)}_{ES}$ (the integrand in~\eqref{eq:SehMultifieldPhiEqPsi}), respectively, 
\begin{equation}
    \Pi_{\Psi} = \frac{\partial \mathscr{L}^{(S)}_{ES}}{\partial \Psi'} = -6\bp^2a^2[\Psi' + \mathcal{H}\Psi], \hspace{1 cm} \Pi_{\varphi i} = \frac{\partial \mathscr{L}^{(S)}_{ES}}{\partial \varphi'^i} = a^2[K_{ij}\varphi'^j+K_{ij,l}\phi'^j\varphi^l - 4 K_{ij}\phi'^j\Psi].
    \label{eq:MultiConjugateMomenta}
\end{equation}
We can invert these equations to find $\Psi'$ and $\varphi'^i$ in terms of $\Psi$ and $\varphi^i$ and their conjugate momenta:
\begin{equation}
    \Psi' = -\frac{\Pi_{\Psi}}{6\bp^2a^2}-\mathcal{H}\Psi, \hspace{1 cm}\varphi'^i = \frac{1}{a^2}K^{ij}\ \Pi_{\varphi j} -K^{ij}K_{jk,l}\phi'^k\varphi^l+ 4\phi'^i \Psi.
    \label{eq:MultiConjugateMomentaInverse}
\end{equation}
Then, the Hamiltonian density is given by:
\bea
    \mathscr{H} &=& \Pi_{\varphi i}\varphi'^i +\Pi_\Psi\Psi' - \mathscr{L}^{(S)}_{ES} =-\frac{\Pi_\Psi^2}{12a^2\bp^2} - \mathcal{H}\ \Pi_\Psi\Psi -a^2\bp^2\ \Psi\vec\nabla^2\Psi +15a^2\bp^2(\mathcal{H}^2-\mathcal{H'})\Psi^2\nonumber \\
    && + \frac{\Pi_{\varphi}^i\Pi_{\varphi i} }{2a^2}-K_{ij,l}\phi'^i\Pi_{\varphi}^j\varphi^l - \frac{a^2}{2} \ K_{ij} \varphi^i\vec\nabla^2\varphi^j+\frac{a^2}{2}K^{ij}K_{jk,l}K_{im,n}\phi'^k\phi'^m\varphi^l\varphi^n \nonumber \\
  && +\frac{a^4}{2}U_{,i,j} \varphi^i\varphi^j+4\Psi\phi'^i\Pi_{\varphi i}-2a^2\Psi K_{mn,p}\phi'^m\phi'^n\varphi^p-2a^4\Psi U_{,i}\ \varphi^i.\label{Hscalar} 
\eea

One can now derive the field equations from these Hamiltonian, but as anticipated one should also add Eq.~\eqref{eq:MultiScalPertSys2}, which is a primary constraint in our approach. In the Hamiltonian formalism this constraint reads
\be C_1 \equiv \Pi_\Psi + 3 a^2K_{ij}\phi'^i \varphi^j = 0.\label{C1ex}\ee
To describe it properly we construct a total Hamiltonian density by adding $C_1$ times a Lagrange multiplier $\lambda$ (a real scalar field):
\begin{equation}
    \mathscr{H}_{\rm tot} = \mathscr{H}\ +\ \lambda C_1\, .\label{Htot} 
\end{equation}

This primary constraint leads to the following secondary constraint, which ensures that $C_1=0$ is  preserved at any time on a solution of the equations of motion\footnote{Note that the partial derivative term $\frac{\partial C_1}{\partial \eta}$ is needed as $C_1$ also  depends on $a(\eta)$ and $\phi(\eta)$.}:
\bea
C_2 &\equiv&  \frac{dC_1}{d\eta} = \frac{\partial C_1}{\partial \eta} + \int d^3y\ \{C_1,\mathscr{H}_{\rm tot}(\vec y)\}=\frac{\partial C_1}{\partial \eta} + \int d^3y\ \{C_1,\mathscr{H}(\vec y)\}  \label{eq:C2H}  \\
    &=&  \mathcal{H}\ \Pi_{\Psi} - \phi'^i\ \Pi_{\varphi i}+ 2a^2\bp^2\ \vec\nabla^2\Psi\notag -6a^2\bp^2(\mathcal{H}^2-\mathcal{H}')\Psi-a^4 U_{,i}\varphi^i+\frac{a^2}{2}K_{mn,j}\phi'^m\phi'^n\varphi^j=0,
\eea
where, given two generic functionals, $A_1$ and $A_2$, of the canonical ``coordinates" $X^i$ and their conjugate momenta $\Pi_i$, the (equal-time\footnote{In writing equal-time Poisson (and, later, Dirac) brackets between fields, we leave the time dependence implicit.}) Poisson brackets are defined by
\begin{equation}
    \{A_1,A_2\} \equiv \sum_i \int d^3 x
    \left(
        \frac{\delta A_1}{\delta X^i(\vec{x})}
        \frac{\delta A_2}{\delta\, \Pi_i(\vec{x})}
        -
        \frac{\delta A_1}{\delta\, \Pi_i(\vec{x})}
        \frac{\delta A_2}{\delta X^i(\vec{x})}
    \right)
\end{equation}
and in~\eqref{eq:C2H} we have left the spacetime dependence of $C_1$ implicit.
In our case the $X^i$ include the $\varphi^j$ and $\Psi$. In deriving the expression of $C_2$ above we used the background Eq.~\eqref{eq:phi'2MF}.

No additional constraints arise, indeed, the time-derivative of $C_2$, reads (leaving the spacetime dependence of $C_2$ implicit)
\be \frac{dC_2}{d\eta} = \frac{\partial C_2}{\partial \eta} + \int d^3y\ \{C_2,\mathscr{H}_{\rm tot}(\vec y)\} =  \frac{\partial C_2}{\partial \eta} + \int d^3y\ \{C_2,\mathscr{H}(\vec y)\} + \int d^3y\ \lambda(\vec y)\ \{C_2,C_1(\vec y)\}\label{dC2}\ee
and the last term (the only one that depends on the Lagrange multiplier $\lambda$) generically contributes because the  Poisson bracket between the two constraints is not zero:
\begin{equation}
        \{C_2(\vec x),C_1(\vec y)\} = 2 a^2\bp^2\vec\nabla^2_y\ \delta^{(3)}(\vec{x}-\vec{y}) \neq 0, \label{Pc1c2}
\end{equation}
where $\vec\nabla^2_y$ is the Laplacian acting on $\vec{y}$ and again we used the background Eq.~\eqref{eq:phi'2MF}. Therefore, Eq.~\eqref{dC2} is not an independent constraint, but can rather be solved to find $\lambda$. We note that this solution for $\lambda$ should be set to zero on the solution of the equations of motion to ensure that the system of equations that we want,~\eqref{eq:MultiScalPertSys1}-\eqref{eq:MultiScalPertSys3}, is reproduced\footnote{Indeed, note that $\lambda$ gives a generically non-zero contribution $-3a^2\lambda K_{ij}\phi'^j$ to the right-hand side of the equations of motion $\Pi'_{\varphi i} = - \frac{\partial \mathscr{H}_{\rm tot}}{\partial \varphi^i}$
and does not modify the definition of the $\Pi_{\varphi i}$.}. 
 This setting can be achieved by appropriate initial conditions on the $\varphi^i$, $\Psi$ and their conjugate momenta:
  the consistency of these conditions is ensured by the fact that the system of equations that we want is consistent as they come from the full field equations from the action in~\eqref{eq:SehMultifield} (fixing the gauge only after performing the variations with respect to all dynamical variables) and can be obtained by using the field equations  using $\mathscr{H}$ (rather than $\mathscr{H}_{\rm tot}$) as time-evolution generator plus the constraint, Eq.~\eqref{eq:MultiScalPertSys2}.
 
 The constraints we have, $C_1$ and $C_2$, are second-class constraints as the corresponding ``matrix",  
 \begin{equation}
    C_{ij}(\vec{x},\vec{y}) \equiv \left\{ C_i(\vec{x}),C_j(\vec{y}))\right\}  =  2a^2\bp^2\vec\nabla^2_y\ \delta^{(3)}(\vec{x}-\vec{y})\ \epsilon_{ij},
\end{equation}
       is not degenerate. In the expression above $\epsilon_{ij}$ forms the $2\times 2$ antisymmetric matrix with $\epsilon_{21}=1$ (cf.~Eq.~\eqref{Pc1c2}).  The inverse of the ``matrix" $C$ is given by
\begin{equation}
    C^{ij}(\vec{x},\vec{y}) =\frac{G(\vec{x},\vec{y})}{2a^2\bp^2}  \epsilon^{ij},
\label{eq:Cmatrix}
\end{equation}
where $\epsilon^{ij}$ satisfies $\epsilon^{ik}\epsilon_{kj}= \delta^i_{j}$ and $G(\vec{x},\vec{y})$ represents the inverse of the operator $\vec\nabla^2_y\ \delta^{(3)}(\vec{x}-\vec{y})$, i.e.
\begin{equation}
    \int d^3z\ G(\vec{x},\vec{z})\ \vec\nabla^2_z\ \delta^{(3)}(\vec{z}-\vec{y}) = \delta^{(3)}(\vec{x}-\vec{y}).
    \label{eq:LaplacianGreen}
\end{equation}
 For later use, we express this operator in its explicit form with a Fourier transform,
\begin{equation}
 G(\vec{x},\vec{y}) =- \int\frac{d^3q}{(2\pi)^3}\frac{e^{i\vec{q}\cdot(\vec{y}-\vec{x})}}{q^2}.
    \label{eq:Gxy}
\end{equation}
       We note that the ``matrix" $C$ has exactly the same form in single-field and multi-field inflation.
       
\subsection{Dirac brackets and quantization}\label{Dirac brackets and quantization}

In a theory like ours, with only second-class constraints, $C_i$, the canonical quantization procedure is performed by equating the commutator of two generic operators $A_1$ and $A_2$, which depend on the canonical ``coordinates" $X^i$ and their conjugate momenta $\Pi_i$, with the corresponding Dirac (not Poisson) bracket times $i$~\cite{MNtheorem}, as  proposed by Dirac\footnote{For other applications of the Dirac quantization in the context of cosmological perturbations see Refs.~\cite{Langlois:1994ec,Armendariz-Picon:2016dgd,Malkiewicz:2018ohk}.}:
\be [A_1,A_2]=i\{A_1,A_2\}_D\equiv i \{A_1,A_2\} -i\iint d^3z\ d^3w\ \{ A_1, C_i(\vec{z}) \} \, C^{ij}(\vec{z},\vec{w}) \, \{ C_j(\vec{w}), A_2\},
\label{eq:DiracP}
  \ee 
where we assumed that the constraints are local in space, as is the case here. To compute commutators in this context one therefore needs the Poisson brackets between the canonical variables and the constraints, which are given in Appendix~\ref{List of Dirac brackets} together with the expressions for the Dirac brackets.

This non-trivial quantization will allow us to express the creation and annihilation operators for scalars as functionals of the scalar fields, analogously to what we obtained in Eq.~\eqref{eq:alphahexp} for tensor perturbations. To this end, we consider the $2N$-dimensional real vector space $\mathscr{V}$ of solutions of the field equations  in~\eqref{eq:MultiScalPertSys1}-\eqref{eq:MultiScalPertSys3}  in momentum space ($\vec\nabla^2\to-q^2$), and a generic basis in this space, denoted by $\tilde\varphi_{qr}$, where $r= 1,\dots,2N$ labels the different basis vectors. Each of these vectors has components $\tilde\varphi^i_{qr}$, with $i= 1,\dots,N$ labeling the fields $\varphi^i$. It is not necessary to list here the corresponding solutions $\tilde\Psi_{qr}$ for the field $\Psi$ in momentum space, as these can be expressed as a linear combination of the  $\tilde\varphi^i_{qr}$ through Eq.~\eqref{eq:MultiScalPertSys3}. Moreover, the conjugate momenta in momentum space $\tilde\Pi^i_{qr}$ associated with the $\tilde\varphi^i_{qr}$ can then be expressed as a linear combination of the $\tilde\varphi^i_{qr}$ and the time derivative $\tilde\varphi'^i_{qr}$ through the second equation in~\eqref{eq:MultiConjugateMomenta}. Equivalently, one could also express the $\tilde\varphi'^i_{qr}$ as a linear combination of the $\tilde\varphi^i_{qr}$ and the $\tilde\Pi^i_{qr}$.

We define now an antisymmetric bilinear form $W$ acting on $\mathscr{V}\times \mathscr{V}$ as
\begin{equation}
W_{rs} \equiv W(\tilde\varphi_{qr},\tilde\varphi_{qs}) \equiv \tilde\xi^i_{qr}\  \Omega_{ij}\  \tilde\xi^j_{qs},
\label{eq:Symplecticform}
\end{equation}
where $\Omega$ is a non-degenerate antisymmetric matrix, whose precise choice is postponed, and $\tilde\xi_{qr} = \{\tilde\varphi^1_{qr}, ... , \tilde\varphi^N_{qr}, \tilde\Pi^1_{qr}, ... , \tilde\Pi^N_{qr}\}$. Hence the index $i$ on $\tilde\xi^i_{qr}$ and the indices $ij$ on $\Omega_{ij}$  range from $1$ to $2N$ with unit steps.  Note that $\tilde\xi^i_{qr}$ is linear in $\tilde\varphi^i_r$, ensuring the bilinearity of $W$, while the antisymmetry of $\Omega$ clearly implies the antisymmetry of $W$. It can also be shown that $W$ is non-degenerate iff  $\Omega$ is non-degenerate (see Appendix~\ref{appendix:SymplecticForm}); therefore $W$ is what is known as a symplectic form.

Now, according to a theorem on symplectic forms \cite{Lee2013}, it is always possible to choose a basis $\tilde\varphi_{qr}$ such that the matrix $\mathcal{W}$ with elements $W_{rs}$ takes, at some arbitrary reference time $\eta_*$, its canonical form:
 \begin{equation}
   \mathcal{W} = \begin{pmatrix}
0 & I_N \\
- I_N & 0
\end{pmatrix},
    \label{eq:SymplecticformCanonical}
\end{equation}
where $I_N$ is the $N\times N$ identity matrix.
This is equivalent to
\begin{equation}
    W(\tilde\varphi_{qr},\tilde\varphi_{q\,N+s}) =-W(\tilde\varphi_{q\, N+r},\tilde\varphi_{qs}) =  \delta_{rs}, \hspace{1cm}W(\tilde\varphi_{qr},\tilde\varphi_{qs}) =W(\tilde\varphi_{q\, N+r},\tilde\varphi_{q\, N+s}) = 0,
\end{equation}
with $r,s= 1,\ldots,N$. The necessity of fixing a reference time $\eta_*$ comes from the fact that a change of basis in $\mathscr{V}$, such as the one that realizes~\eqref{eq:SymplecticformCanonical}, must of course   be time independent. 
Starting from these $\tilde\varphi_{qr}$, we construct a complex basis of solutions as follows:
\begin{equation}
    \varphi^i_{qr} = \frac{\tilde\varphi^i_{qr} - i \tilde\varphi^i_{q\, N+r}}{\sqrt{2}},\hspace{1.2cm}\varphi^{*i}_{qr} = \frac{\tilde\varphi^i_{qr} + i \tilde\varphi^i_{q\,  N+r}}{\sqrt{2}},
\end{equation}
with $r= 1,\ldots,N$.
With this choice, we have in the new basis\footnote{Although not directly needed for this work, it is perhaps worth noting that this basis is not unique. Indeed, one can still perform a transformation that does not mix the two subspaces of solutions $\varphi^{i}_{qr}$ and $\varphi^{*i}_{qr}$,
\begin{equation}
    \tilde{\varphi}^i_r = U_{rs}\varphi^i_s\hspace{1.2cm}\tilde{\varphi}^{*i}_r = U_{rs}^*\varphi^i_s,
\end{equation}
and Conditions~\eqref{eq:Wcond1}-\eqref{eq:Wcond2} are preserved if the matrix $U$ is unitary.} at $\eta=\eta_*$:
\begin{align}
 \hspace{-0.505cm}   W(\varphi_{qr}, \varphi^*_{qs}) = \frac{(W(\tilde\varphi_{qr},\tilde\varphi_{qs}) + i W(\tilde\varphi_{qr},\tilde\varphi_{q\, N+s})-iW(\tilde\varphi_{q\, N+r},\tilde\varphi_{qs}) + W(\tilde\varphi_{q\, N+r},\tilde\varphi_{q\, N+s})}{2} = i\delta_{rs}\hspace{-0.3cm} 
    \label{eq:Wcond1}
\end{align}
and similarly
\begin{equation}
    W(\varphi_{qr}, \varphi_{qs}) = W(\varphi^*_{qr}, \varphi^*_{qs}) = 0.
        \label{eq:Wcond2}
\end{equation}

We can now decompose the  Hermitian fields $\varphi^i$ in creation and annihilation operators, $\alpha_r^\dagger(\vec q)$ and $\alpha_r(\vec q)$, respectively,  using this basis:
\begin{align}
\varphi^i(\vec{x},\eta) &= \sum_{r\ =\ 1}^N\int d^3q \left[ \varphi^i_{qr}(\eta) e^{i\vec{q}\cdot\vec{x}} \alpha_r(\vec{q}) + \varphi_{qr}^{*i}(\eta) e^{-i\vec{q}\cdot\vec{x}} \alpha_r^\dagger(\vec{q}) \right],
\label{eq:MultiVarphiExp}
\end{align}
Also, by linearity, $\Pi^i$ must involve the same creation and annihilation operators as $\varphi^i$:
\begin{align}
\Pi_\varphi^i(\vec{x},\eta) &= \sum_{r\ =\ 1}^N\int d^3q \left[ \Pi^i_{qr}(\eta) e^{i\vec{q}\cdot\vec{x}} \alpha_r(\vec{q}) + \Pi_{qr}^{*i}(\eta) e^{-i\vec{q}\cdot\vec{x}} \alpha_r^\dagger(\vec{q}) \right],
\label{eq:MultiPiExp}
\end{align}
where the $\Pi^i_{qr}$ are the conjugate momenta  associated with the  $\varphi^i_{qr}$. 

The symplectic form $W$ we have constructed above can now be used to invert these relations and express the $\alpha_r(\vec q)$ as functionals of the fields. To show this we first go to momentum space, 
\begin{align}
    &\varphi^i(\vec{q}) \equiv \varphi^i_{qr}\ \alpha_r(\vec{q})+\varphi^{*i}_{qr}\ \alpha_r^\dagger(-\vec{q}), \\
    &\Pi^{i}_\varphi(\vec{q}) \equiv \Pi^i_{qr}\ \alpha_r(\vec{q})+\Pi^{*i}_{qr}\ \alpha_r^\dagger(-\vec{q}) 
\end{align}
or, more compactly, using $\xi_{qr} \equiv \{\varphi^1_{qr}, ... , \varphi^N_{qr}, \Pi^1_{qr}, ... , \Pi^N_{qr}\}$
\begin{equation}
\xi^i(\vec{q}) \equiv \xi^i_{qr}\ \alpha_r(\vec{q})+\xi^{*i}_{qr}\ \alpha_r^\dagger(-\vec{q}),
\label{eq:xiFourierExp}
\end{equation}
where the index $i$ on $\xi^i_{qr}$ ranges from $1$ to $2N$ with unit steps.
Contracting now both sides with $\xi^{*j }_{qs}\  \Omega_{ji}$,
 \begin{equation}
 \xi^{*j }_{qs}\  \Omega_{ji} \xi^i(\vec{q}) = \xi^{*j }_{qs}\  \Omega_{ji} \xi^i_{qr} \alpha_r(\vec{q}) +  \xi^{*j }_{qs}\  \Omega_{ji}\xi^{*i}_{qr} \alpha_r^\dagger(-\vec{q}),
 \end{equation}
setting $\eta=\eta_*$ and using the property in~\eqref{eq:Wcond1} of the $\varphi_{qr}$ basis, we obtain
\begin{equation}
\boxed{\alpha_r(\vec{q}) =  i\xi^{*i}_{qr}\  \Omega_{ij} \xi^j(\vec{q}).} \label{aExFF}
\end{equation}
Going back from momentum to coordinate space (expressing the $\xi^j$ in terms of the $\varphi^i$ and $\Pi^i_\varphi$ by inverting~\eqref{eq:MultiVarphiExp}-\eqref{eq:MultiPiExp}), the last equation precisely gives us the $\alpha_r(\vec{q})$ as functionals of the fields, $\varphi^i$ and $\Pi^i_\varphi$.

We can now uniquely determine the commutator between annihilation and creation operators:
\begin{equation}
\left[ \alpha_r(\vec{q}),\alpha^\dagger_s(\vec{k}) \right] =  \xi^{*m}_{qr}\  \Omega_{mi}\ \xi^{ n}_{ks}\  \Omega_{nj}\left[ \xi^i(\vec{q}), \xi^j(-\vec{k}) \right], \label{eq:alphaalphadaggerxiixij}
\end{equation}
where we used $\alpha^\dagger_s(\vec{q}) =- i\ \xi^{n}_{qs}\  \Omega_{nj} \xi^{ j}(-\vec{q})$, which follows from~\eqref{aExFF}. We can decompose the commutator $\left[ \xi^i(\vec{q}), \xi^j(-\vec{k}) \right]$ on the right-hand side of~(\ref{eq:alphaalphadaggerxiixij}) as
\begin{equation} 
\begin{pmatrix}
\left[ \varphi^i(\vec{q}), \varphi^j(-\vec{k}) \right] & \left[ \varphi^i(\vec{q}), \Pi_\varphi^j(-\vec{k}) \right]\\[6pt]
\left[ \Pi_\varphi^i(\vec{q}), \varphi^j(-\vec{k}) \right]&\left[ \Pi_\varphi^i(\vec{q}), \Pi_\varphi^j(-\vec{k}) \right]
\end{pmatrix}.
\label{eq:xiixijCommutator}
\end{equation}
The corresponding commutators in coordinate space are given by the Dirac brackets in Appendix~\ref{List of Dirac brackets} multiplied by $i$, Eq.~\eqref{eq:DiracP}. So, for instance, one obtains $\left[ \varphi^i(\vec{q}), \varphi^j(-\vec{k}) \right]=0$  and 
\bea
 \left[ \varphi^i(\vec{q}), \Pi_\varphi^j(-\vec{k}) \right] &=& \int \frac{d^3x}{(2\pi)^3}\frac{d^3y}{(2\pi)^3} e^{-i\vec{q}\cdot\vec{x} + i\vec{k}\cdot\vec{y}}  \left[ \varphi^i(\vec{x}), \Pi_\varphi^j(\vec{y}) \right]  \nonumber\\ &=&
 i\ \int \frac{d^3x}{(2\pi)^3}\frac{d^3y}{(2\pi)^3} e^{-i\vec{q}\cdot\vec{x} + i\vec{k}\cdot\vec{y}} \left (K^{ij}\ \delta^{(3)}(\vec{x}-\vec{y}) -\frac{3}{2\bp^2}\phi'^i \phi'^j\  G(\vec{x},\vec{y})\right)  \nonumber\\
 &=& \frac{i}{(2\pi)^3}\left(K^{ij}+ \frac{3}{2\bp^2q^2} \phi'^i \phi'^j\right)\ \delta^{(3)}(\vec{q}-\vec{k})\equiv i A^{ij}\ \delta^{(3)}(\vec{q}-\vec{k}), \label{DefA}
\eea
where all quantities are evaluated at $\eta=\eta_*$ and we used the explicit expression of $G(\vec{x},\vec{y})$ in~(\ref{eq:Gxy}). 
Similarly, we obtain $\left[ \Pi^i(\vec{q}), \Pi^j(-\vec{k}) \right] = i B^{ij} \delta^{(3)}(\vec{q}-\vec{k})$, where $B^{ij}$ is antisymmetric, $B^{ij}=-B^{ji}$, unlike $A^{ij}$ defined in~(\ref{DefA}), which is instead symmetric, $A^{ij}= A^{ji}$. Therefore, the commutator $\left[ \xi^i(\vec{q}), \xi^j(-\vec{k}) \right]$ on the right-hand side of~(\ref{eq:alphaalphadaggerxiixij}) has the structure 
\begin{equation}
\left[ \xi^i(\vec{q}), \xi^j(-\vec{k}) \right] = 
iM^{ij}  \delta^{(3)}(\vec{q}-\vec{k}),
\end{equation}
where the matrix $M$ is 
\begin{equation}
M = \begin{pmatrix}
 0 &  A\\
-A & B\ 
\end{pmatrix}.
\end{equation}
This matrix is antisymmetric (as $A$ is symmetric and $B$ is antisymmetric) and its determinant equals $(\det A)^2$. Since $A$ is non degenerate\footnote{We observe that $A^i_j$ has $N-1$ eigenvectors, $v^i_{(n)}$, with $n= 1,\dots,N-1$, corresponding to directions orthogonal to $\phi'^i$: the orthogonality condition reads $K_{ij} \phi'^i v^j_n=0$.
The corresponding eigenvalues are all equal to $1/(2\pi)^3\neq0$:
\begin{equation}
    \frac{1}{(2\pi)^3}\left(\delta^i_j+\frac{3}{2\bp^2\ q^2}\phi'^iK_{jl}\phi'^l\right)v^j_{(n)} =  \frac{1}{(2\pi)^3} v^i_{(n)}.
\end{equation}
The last eigenvector, which is proportional to $\phi'^i$ and we call $v^i_{\parallel}$,  satisfies:
\begin{equation}
    \frac{1}{(2\pi)^3}\left(\delta^i_j+\frac{3}{2\bp^2\ q^2}\phi'^iK_{jl}\phi'^l\right)v^j_{\parallel} =  \frac{1}{(2\pi)^3}\left(1+\frac{3K_{jl}\phi'^j\phi'^l}{2\bp^2\ q^2}\right)\ v^i_\parallel.
\end{equation}
So the last eigenvalue is also non zero, and thus $A$ is non degenerate.}, $M$ is also non degenerate. Since all we have assumed so far about $\Omega$ is that it is an antisymmetric, non-degenerate matrix, we can now take it to be the inverse of $-M$. With this choice, using~\eqref{eq:Wcond1} in~\eqref{eq:alphaalphadaggerxiixij}, we obtain   \begin{equation}
\left[ \alpha_r(\vec{q}),\alpha^\dagger_s(\vec{k}) \right] = \delta_{rs}. \label{usualComm}
\end{equation}
Regarding $\left[ \alpha_r(\vec{q}),\alpha_s(\vec{k}) \right]$, the procedure is analogous, and leads to
\begin{equation}
\left[ \alpha_r(\vec{q}),\alpha_s(\vec{k})\right] = -\xi^{*m}_{qr}\  \Omega_{mn}\ \xi^{*n}_{qs}\ i\  \delta^{(3)}(\vec{q}+\vec{k}),
\end{equation}
which, due to condition \eqref{eq:Wcond2}, vanishes:
\begin{equation}
     \Big[\alpha_r(\vec{q}),\alpha_s(\vec{k})\Big] = 0.\label{usualComm2}
\end{equation}
Therefore, using the Dirac quantization of constrained systems, also the Hilbert space and amplitude of quantum fluctuations of scalars can be determined without relying on the sub-horizon limit. They are determined, respectively, by the commutation rules in~\eqref{usualComm} and~\eqref{usualComm2} and by a choice of basis that satisfies~\eqref{eq:SymplecticformCanonical}.

On the other hand, observable inflationary power spectra are computed at horizon exit  $q/a=H$. As a result, all that is needed to reconstruct the Hilbert space and amplitude of quantum fluctuations is to consider times where  $q/a$ was of order $H$. Therefore, neglecting higher-derivative terms produces uncertainties of relative order $(H/\Lambda)^n$ where $n$ is the minimal number of extra derivatives that are neglected. The Planck collaboration determined the upper bound $H < 2.7\cdot 10^{-5}\bp \approx 6.6 \cdot 10^{13}$\,GeV~\cite{Ade:2015lrj} at a reference pivot scale, so for $\Lambda\gg6.6 \cdot 10^{13}$\,GeV these uncertainties are always small. In a purely bosonic relativistic theory, such as the generic two-derivative inflationary model with Action~(\ref{eq:SehMultifield}), the higher-derivative terms with the smallest number of derivatives that one can add are four-derivative terms, so $n=2$ and one has a relative suppression of order $(H/\Lambda)^2$. The case $n=1$ may occur in the presence of fermions, which are, however, typically neglected during the inflationary stage of the universe. The ratio $H/\Lambda$ is therefore the key quantity that controls the size of the higher-derivative corrections in an EFT. 

\section{Examples and applications}\label{Applications}

Here we illustrate how to use the results of Secs.~\ref{Tensor perturbations} and~\ref{Scalar perturbations}
 through some examples and applications. 

Let us start by mentioning possible UV completions of Einstein gravity (see~\cite{Buoninfante:2024yth} for a more general overview) and the possible value of the cutoff $\Lambda$ of the low-energy EFT constructed from these completions. 

One class of UV completions of Einstein gravity is the softened-gravity scenario proposed and studied in~\cite{Giudice:2014tma,Salvio:2016vxi}. In this class the gravitational interactions mimic those of Einstein gravity up to an energy scale $\Lambda_{\rm soft}\ll \bp$, but get softened at energies of order or above $\Lambda_{\rm soft}$. The main motivation for this behavior is the Higgs mass naturalness problem:  in the softened-gravity scenario this problem can be solved if $\Lambda_{\rm soft}$ is no larger than $10^{11}$\,GeV. In any of these theories the low-energy EFT constructed by integrating out the extra degrees of freedom responsible for the softening must have a cutoff $\Lambda<\Lambda_{\rm soft}$. If this is the way the Higgs mass naturalness problem is solved the only inflationary models that can be consistently described in such a low-energy EFT are low-scale models with $H\ll 10^{11}$\,GeV. Examples of models of this type include the curvaton models (see e.g.~\cite{Dimopoulos:2004yb,BuenoSanchez:2007jxm} and~\cite{Torrado:2017qtr,Byrnes:2025kit} for more recent analysis).
Inflation may still be viable at energies exceeding $\Lambda_{\rm soft}$ if the theory includes the extra degrees of freedom responsible for the softening of gravity. Examples of possible embeddings of the softened-gravity idea may be string theories with low string scale~\cite{Antoniadis:1990ew} or quadratic gravity with large values of the coefficients of the quadratic-in-curvature terms~\cite{Salvio:2014soa,Salvio:2017qkx,Salvio:2020axm,Salvio:2024joi}. 

Einstein gravity can also be completed if the coefficients of the (infinite) possible terms of the gravitational EFT run with energies in a way that in the UV they reach a finite-dimensional surface (asymptotic safety)~\cite{WeinbergAS,Eichhorn:2018yfc}. In this case, which could also be realized in quadratic gravity~\cite{Benedetti:2009rx,Falls:2020qhj}, the gravitational EFT can still be treated perturbatively up to a cutoff scale $\Lambda$ which can be as large as $\bp$. 

\vspace{0.3cm}

We now analyze some well-known and successful inflationary scenarios and models to estimate the uncertainties due to the higher-derivative terms that are not included in the action in~(\ref{eq:SehMultifield}), $(H/\Lambda)^2$. We will keep in mind that the maximal value of $\Lambda$ is $\bp$.

In the ``new inflation" scenario, where inflation is driven by a slowly-rolling inflaton $\chi$ (the current paradigm for the inflationary epoch), the curvature power spectrum $P_R$ (at horizon exit) is given by
\be P_R\approx \frac{U/ \epsilon}{24\pi^2 \bp^4} \label{PRsingle}\ee
in the slow-roll approximation, where 
\be \epsilon \equiv  \frac{\bp^2}{2} \left(\frac{1}{U} \frac{dU}{d\chi}\right)^2 \label{epsilon}\ee
is the first slow-roll parameter and we have assumed $\chi$ to be canonically normalized. At the inflationary scales these quantities are evaluated at an appropriate number of e-folds $N_e$
  \be N_e(\chi)  = \frac1{\bp^2}  \int^\chi_{\chi_{\rm end}} d\chi'  \,  U\left(\frac{dU}{d\chi'}\right)^{-1} \label{Ne} \ee 
 before the end of inflation, which can be defined through the field value $\chi_{\rm end}$ such that $\epsilon(\chi_{\rm end})=1$. This appropriate number is around $60$ for many relevant models. 

The observed value of $P_R$ is $(2.10 \pm 0.03) \cdot 10^{-9}$~\cite{Ade:2015lrj}. 
 The slow-roll approximation also gives 
\be H^2\approx \frac{U}{3\bp^2}.  \label{H2vsU}\ee
 Therefore, using~\eqref{PRsingle} and~\eqref{H2vsU}
\be \left(\frac{H}{\Lambda}\right)^2 \approx  8\pi^2 P_R \ \epsilon \  \frac{\bp^2}{\Lambda^2} \approx 1.7 \cdot 10^{-7} \epsilon \  \frac{\bp^2}{\Lambda^2}.  \label{HDgeps}\ee 
This result relates the size of the corrections due to higher-derivative terms to the size of $\epsilon$ for a given $\Lambda$.
We deduce that neglecting higher-derivative terms in the EFT is always better than the slow-roll approximation if\footnote{Using the results in~\cite{Stewart:1993bc}, one finds that in general the leading-order slow-roll approximation has an uncertainty of relative order $\gtrsim\max(\epsilon,|\eta|)$.} 
\be \Lambda  \gtrsim 4.1 \cdot 10^{-4}\bp. \label{boundG}\ee
This bound can be significantly relaxed in slow-roll models where the second slow-roll parameter
\be \eta \equiv \frac{\bp^2}{U} \frac{d^2U}{d\chi^2} \label{eta}\ee
has an absolute value significantly larger than that of $\epsilon$ at the inflationary scales and so it is not  $\epsilon$ that gives the leading uncertainty in the slow-roll approximation. For those models neglecting higher-derivative terms in the EFT is  better than the slow-roll approximation if 
\be \Lambda  \gtrsim 4.1 \cdot 10^{-4}  \sqrt{\frac{\epsilon}{|\eta|}}  \, \bp,  \qquad (|\eta|>\epsilon). \label{LamEta}\ee
We will provide some examples of models of this type in Secs.~\ref{Starobinsky}-\ref{Hilltop}.

Moreover, it is worth mentioning that Eq.~\eqref{H2vsU} illustrates how a small $(H/\Lambda)^2$ also guarantees a sub-Planckian value of the inflationary potential density $U$ as $\Lambda\leq\bp$. This is a necessary condition to keep quantum gravitational effects small. 

Note also that these estimates do not significantly change in realistic multi-field inflationary models because of the stringent Planck bounds on isocurvature perturbations~\cite{Ade:2015lrj}. It may however be worth pointing out that the estimates above can be easily extended to the generic multi-field case by substituting~\eqref{PRsingle} with
\be P_R\approx \frac{2H^2}{\pi^2 r \bp^2}, \label{PRmulti}\ee
where $r$ is the tensor-to-scalar ratio. This relation holds for an arbitrary number of inflatons and replaces~\eqref{HDgeps} with
\be \left(\frac{H}{\Lambda}\right)^2 \approx  \frac{\pi^2}{2} P_R \ r \  \frac{\bp^2}{\Lambda^2} \approx 1.0 \cdot 10^{-8}\,   r \  \frac{\bp^2}{\Lambda^2}.  \label{HDr}\ee 
Therefore, in an EFT with a given cutoff $\Lambda$ the
size of the corrections due to higher-derivative terms are set by $r$.

We now consider some relevant ``new inflation" models in turn.

\subsection{Starobinsky inflation}\label{Starobinsky}

One of the most famous and successful inflationary models was presented by Starobinsky in Ref.~\cite{Starobinsky:1980te}, a pioneering work on inflation. In~\cite{Starobinsky:1980te} Starobinsky considered  quadratic gravity. The $R^2$ term in the Jordan frame corresponds to a scalar field $\phi_S$, called the scalaron, in the Einstein frame (see~\cite{Salvio:2018crh} for a review). 

The inflationary potential density in the scalaron direction reads   
\be U(\phi_S) =   \frac{3f_0^2 \bp^4}{8}\left(1-e^{-2\phi_S/\sqrt{6}\bp}\right)^2\ee
for a canonically-normalized scalar field ($K_{ij}$ just equals one along the $\phi_S$ directions). The positive parameter $f_0^2$ corresponds to the coefficient of the $R^2$ term in the Jordan frame. The Starobinsky model is a particular case of the general class of models described by the action in~(\ref{eq:SehMultifield}). In Fig.~\ref{SR} we provide an estimate of the size of the corrections due to higher-derivative terms by plotting $\epsilon$ as a function of $N_e$ in a phenomenologically relevant interval. Indeed, $\epsilon$ gives $(H/\Lambda)^2$ via Eq.~\eqref{HDgeps}. 
The maximal value of $\Lambda$ in the Starobinsky model is $\sim \bp$, the energy scale at which quantum gravitational effects become large.

\subsection{Higgs inflation}\label{Higgs}

The most minimal choice for the inflaton is the Higgs field $\phi_H$, since this can be realized within the Standard Model~\cite{Bezrukov:2007ep}. One obtains a viable setup if the Higgs field features a non-minimal coupling such that in the Einstein frame the inflationary potential density in the Higgs direction reads  
\be U(\phi_H) =   \frac{\lambda_H \phi_H^4}{4\Omega^4(\phi_H)}\ee
with $K_{ij}$ along the $\phi_H$ directions given by
\be K_{HH}(\phi_H) = \frac{\Omega^2(\phi_H)+6\xi^2\phi_H^2/\bp^2}{\Omega^4(\phi_H)}\ee
and $\Omega^2(\phi_H)$ is defined by
\be \Omega^2(\phi_H)\equiv1+\xi\phi_H^2/\bp^2. \ee
Here $\lambda_H$ is the Higgs quartic coupling and $\xi$ is a parameter giving the strength of the non-minimal coupling. The relation between $\phi_H$ and the canonically-normalized scalar field $\chi$ is obtained by solving the differential equation
\be \frac{d\phi_H}{d\chi} = \frac1{\sqrt{K_{HH}}}\ee 
and one can express $\epsilon$, $\eta$ and $N_e$ in Eqs.~\eqref{epsilon},~\eqref{eta} and~\eqref{Ne} in terms of $\phi_H$, rather than $\chi$, using the chain rule for the field derivatives.

The expressions above do not take quantum corrections into account and, comparing this model with experimental data and observations, lead to a value of $\xi$ of order $10^4$~\cite{Bezrukov:2007ep}.  It was pointed out that this large value leads to a violation of perturbative unitarity at some high-energy scale~\cite{crit,Burgess:2010zq}. Once the inflationary background fields are taken into account, however, one can show~\cite{Bezrukov:2010jz} that such energy is background-field dependent and turns out to be parametrically higher than all relevant scales during the history of the Universe (some additional assumptions on the underlying ultraviolet  complete theory are necessary to reach this conclusion~\cite{Burgess:2010zq,Bezrukov:2010jz, Bezrukov:2012sa}). Nevertheless, it remains interesting to study the sensitivity of this inflationary model to higher-derivative terms in the EFT. As shown in~\cite{Bezrukov:2010jz}, at the inflationary scales, the background-dependent cutoff in the Einstein frame we are considering reduces to $\Lambda \approx \bp$. 

Taking quantum corrections into account, the value of $\xi$ can be reduced~\cite{Hamada:2014wna,Hamada:2014iga,Bezrukov:2014bra}, partially alleviating the perturbative-unitarity issue~\cite{Salvio:2015kka,Salvio:2017oyf}. In order to be conservative we therefore stick to the classical analysis of~\cite{Bezrukov:2007ep}, which leads to the largest value of $\xi$.  Another way of alleviating the perturbative-unitarity issue is to combine Higgs inflation with Starobinsky inflation~\cite{Gorbunov:2018llf}.

We find that the typical size $(H/\Lambda)^2$ of the higher-derivative corrections is practically indistinguishable from that of the Starobinsky model, as shown in Fig.~\ref{SR}.

\subsection{Natural inflation}\label{Natural}

Another well-motivated inflaton is a pseudo-Nambu-Goldstone boson (PNGB)~\cite{Freese:1990rb,Adams:1992bn}.  Indeed, Goldstone's theorem protects the mass (actually the full potential) of a Nambu-Goldstone boson (NGB) in such a way that the required potential flatness 
becomes natural. Eventually, explicit symmetry-breaking terms (which turn the NGB into a PNGB and introduce a potential slope) are necessary to terminate inflation, but these terms can be small, preserving the naturalness of PNGB-driven inflation. This scenario is consequently also referred to as natural inflation. The natural inflaton $\phi_N$ has a periodic potential  that can be derived from UV complete (e.g.~QCD-like) field theories~\cite{Adams:1992bn}. In the absence of non-minimal couplings, natural inflation with a cosine potential  features an unacceptably large value of $r$, but a non-minimal coupling, however, can cure this situation\footnote{See also Refs.~\cite{Ferreira:2018nav,Simeon:2020lkd} for previous studies using older observational data.}~\cite{Salvio:2023cry}. 

Let us then consider the latter variant in the following discussion. In the Einstein frame the inflationary potential density in the $\phi_N$ direction can be taken to be
\be U(\phi_N) =   \frac{V(\phi_N)}{F^2(\phi_N)},\ee
where 
\begin{equation}
    V(\phi_N) \equiv \Lambda_N^4 \left[ 1 + \cos\left(\frac{\phi_N}{f}\right) \right], \qquad F(\phi_N) \equiv 1 + \alpha\, \left[ 1 + \cos\left( \frac{\phi_N}{f} \right)\right],
    \label{eq:V(phi)}
\end{equation}
$\Lambda_N$ and $f$ are two energy scales and $\alpha$ is a real parameter that must satisfy $\alpha>-1/2$ in order for the effective Planck mass in the Jordan frame  to be real for all $\phi_N$. A microscopic origin of $V(\phi_N)$ and $F(\phi_N)$ in terms of a fundamental QCD-like field theory was provided in Ref.~\cite{Salvio:2021lka}.

As in Higgs inflation, $K_{ij}$ along the $\phi_N$ directions is non trivial and given by
\be K_{NN}(\phi_N) =\frac{2F(\phi_N)+3\bp^2 (\frac{dF}{d\phi_N}(\phi_N))^2}{2F^2(\phi_N)} \ee 
and the relation between $\phi_N$ and the canonically-normalized scalar field $\chi$ is given by
\be \frac{d\phi_N}{d\chi} = \frac1{\sqrt{K_{NN}}}. \ee
The agreement with observations in this model requires trans-Planckian values of $f$. This does not necessarily invalidate the EFT treatment of gravity as long as the energy density is much less than the Planck energy density. This condition is automatically satisfied matching the small observed value of $P_R$ with an appropriate choice of $\Lambda_N\ll \bp$. Indeed,  in some UV completions a trans-Planckian $f$ can be consistent~\cite{Salvio:2019wcp,Salvio:2022mld}. It is worth pointing out, however, that the need of trans-Planckian $f$ can be avoided in other variants of natural inflation~\cite{Racioppi:2024zva,Racioppi:2025pim,Kraiko:2026nas}. Note also that, for moderate values of $\alpha$, this model does not suffer from the perturbative unitarity issue of Higgs inflation and the cutoff can be as large as $\bp$.

In Fig.~\ref{SR} we show the typical size $(H/\Lambda)^2$ of the higher-derivative corrections in natural inflation for $f=6\bp$ and $\alpha=0.2$ (a benchmark value in agreement with the observational data of \cite{Ade:2015lrj,BICEP:2021xfz} for some of the $N_e$ considered in Fig.~\ref{SR}, see Ref.~\cite{Salvio:2023cry}). The prediction deviates from the  Starobinsky/Higgs one.

\subsection{Hilltop inflation}\label{Hilltop}

Another popular inflationary model is the hilltop one, where the inflaton $\phi_h$ slowly rolls down the potential, like a ball rolling down from the top of a hill~\cite{Boubekeur:2005zm}. Here we consider a variant of hilltop inflation where the potential density along the $\phi_h$ direction reads~\cite{Salvio:2019wcp} (see also Ref.~\cite{Linde:2011nh} and~\cite{Barvinsky:2015qea} for another type of hilltop inflation)
\be U(\phi_h) = \frac{\lambda_h}{4} \cosh ^4\left(\frac{\phi_h}{\sqrt{6} \bp}\right) \left(v^2-6 \bp^2 \tanh ^2\left(\frac{\phi_h}{\sqrt{6} \bp}\right)\right)^2. \ee 
This potential is obtained by starting from a conformally non-minimally coupled scalar field with a standard quartic mexican-hat potential,  moving to the Einstein frame and canonically normalizing the inflaton. An advantage of this model is that it can be studied analytically. The parameters $\lambda_h$ and $v$ correspond, respectively, to the quartic coupling and  the energy scale of spontaneous symmetry breaking in the mexican-hat potential.

\begin{figure}[t]
    \centering
    \includegraphics[width=0.46\textwidth]{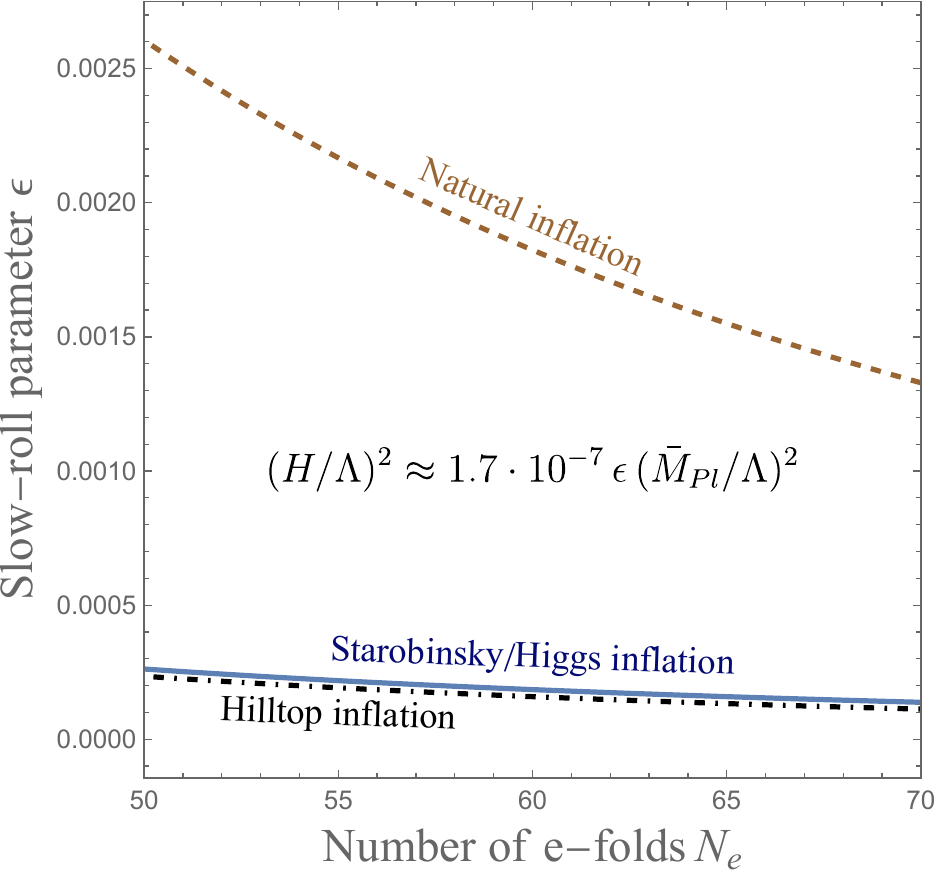} \qquad  \includegraphics[width=0.46\textwidth]{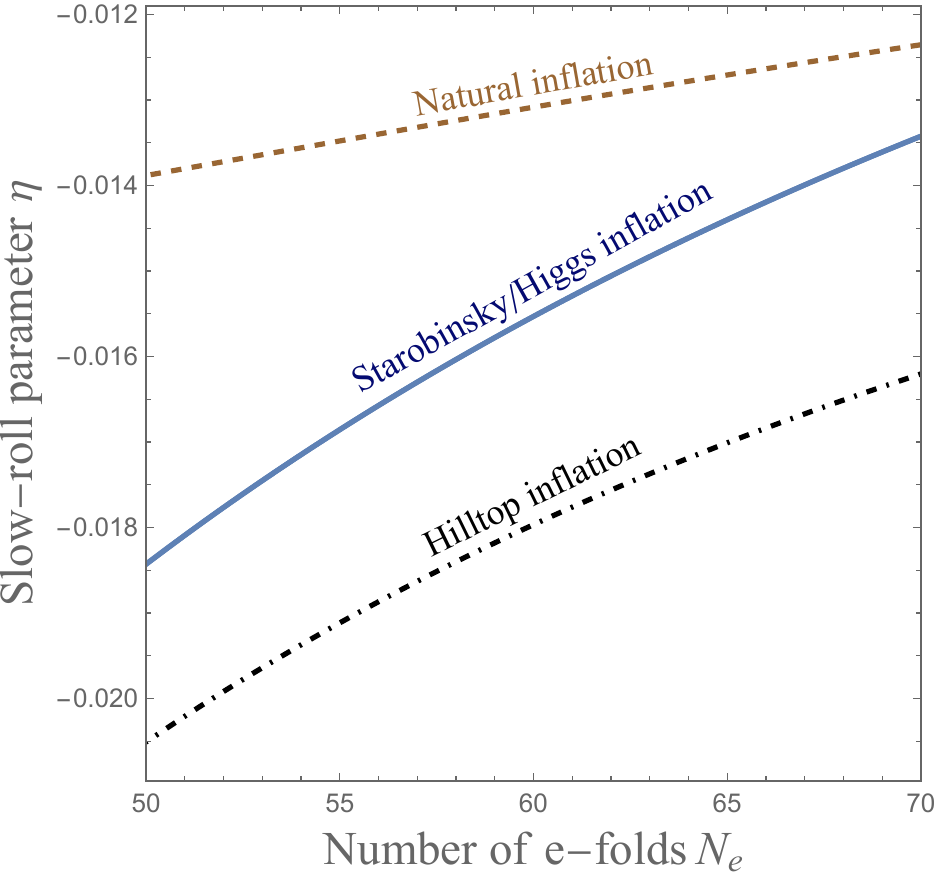}
    \caption{\em {\bf Left:} The relative size $(H/\Lambda)^2$ of the corrections due to higher-derivative terms in the EFT is given in terms of the first slow-roll parameter $\epsilon$ for some famous and successful inflationary models; for these models the maximal value of the cutoff $\Lambda$ is $\bp$.
    {\bf Right:} the second slow-roll parameter $\eta$, which is significantly larger than $\epsilon$ for these models, so the bound in~(\ref{LamEta}) holds. }
    \label{SR}
\end{figure}

In Fig.~\ref{SR} we show the typical size $(H/\Lambda)^2$ of the higher-derivative corrections in hilltop inflation for $v=2.43\bp$ (a benchmark value in agreement with the observational data of \cite{Ade:2015lrj,BICEP:2021xfz} for some of the $N_e$ considered in Fig.~\ref{SR}, see Ref.~\cite{Salvio:2019wcp}). The prediction deviates from both the  Starobinsky/Higgs and the natural-inflation ones.

\section{Summary, discussion and conclusions}\label{conclusions}

The sub-horizon limit simplifies the analysis of cosmological perturbations but obscures the conditions for the validity of an inflationary EFT. It is therefore important to avoid taking this limit in order to identify genuine trans-Planckian problems in inflationary models. 

Motivated by this situation, we have performed the quantization of cosmological perturbations in the general relativistic low-energy (two-derivative) EFT of (multi-)field inflation, avoiding using the sub-horizon limit. We have analysed all types of perturbations and in all cases the general logic to avoid the sub-horizon limit was the following: 
\begin{itemize}
\item One can always express, at an arbitrary time, the annihilation operators $a_i$ as functionals of the fields, see Eqs.~\eqref{eq:alphahexp} and~\eqref{aExFF}.
\item On the other hand, the fields are quantized through canonical equal-time commutators (at the arbitrary time). 
\item Since the canonical commutators are non-homogeneous operator relations, one then fixes the amplitude of quantum fluctuations and find the usual commutation relations $[a_i,a_j]=0$ and $[a_i,a^\dagger_j]=\delta_{ij}$, which determine the Hilbert space.
\end{itemize}

Among the various types of perturbations, the multi-scalar sector is the most complex one as it features constraints and mixings, if one does not use the sub-horizon limit. Specifically this sector presents one primary constraint and one secondary constraint, which are of second-class. To handle this situation we have thus used Dirac quantization of constrained systems and determined the equal-time commutators for all pairs of fields. This allowed us to reconstruct the full Hilbert space and amplitude of quantum fluctuations without relying on the sub-horizon limit.

Therefore, the two-derivative inflationary EFT we considered is affected by possible higher-derivative corrections, but these are of relative order $(H/\Lambda)^2$
 and thus small if $H\ll  \Lambda$: 
 if the cutoff $\Lambda$ is sufficiently larger than the inflationary Hubble rate $H$, neglecting the higher-derivative terms is a good approximation at a given level of accuracy. In some EFTs, such as those obtained from softened gravity motivated by the Higgs-mass naturalness problem, the  cutoff $\Lambda$ has to be much smaller than $\bp$. But other scenarios can have $\Lambda\sim\bp$.

Although  the slow-roll approximation was not used to find  the general results mentioned above, we have subsequently considered in some detail slow-roll inflation, the current paradigm for the inflationary epoch. In this context we have also found a way to estimate the size of the higher-derivative corrections in terms of $\Lambda$  and the tensor-to-scalar ratio $r$ (or, for single-field inflationary models, in terms of $\Lambda$  and the first slow-roll parameter $\epsilon$). This allowed us to find a lower bound on $\Lambda$, Eqs.~\eqref{boundG} and~\eqref{LamEta}, which ensures that the uncertainty due to higher-derivative corrections is smaller than that of the slow-roll approximation.

These results have been applied here to a number of famous and successful inflationary models with a finite cutoff: Higgs inflation, scalaron inflation, natural inflation and hilltop inflation. We have shown in Fig.~\ref{SR} that the size of the higher-derivative corrections is model dependent for a given $\Lambda$ and displayed its values for the above-mentioned models.

Our paper definitely leaves some interesting questions for future research. While we have been able to fix the Hilbert space and amplitude of quantum fluctuations at an arbitrary time (therefore without using the sub-horizon limit),
we have not constrained  the inflationary quantum state. Such state is usually chosen to be the Bunch-Davies vacuum, but  other choices remain possible. This does not directly appear to be a lack of predictivity of the  EFT because even the use of the sub-horizon limit would not fix the Bunch-Davies vacuum as the only possible inflationary state: this is because an excited state can in principle be sufficiently long lived to be viable~\cite{Armendariz-Picon:2006saa}. However, not having found any theoretical preference for the Bunch-Davies vacuum or any other state even in the analysis performed at arbitrary time (not using the sub-horizon limit) does motivate the analysis of different choices for the inflationary state. In the literature there are already some studies. For example, Ref.~\cite{Chung:2003wn} also considered the so-called adiabatic vacuum, but found that the uncertainty due to the choice of the inflationary state is extremely small,  in general likely negligible compared to other uncertainties. A recent paper~\cite{Wood-Saanaoui:2026qni} analysed the so-called  $\alpha$-vacuum, but found that within the regime of validity of the EFT, $H\ll\Lambda$, the predictions reduce to those of the standard Bunch-Davies vacuum. A general inflationary pure state was previously studied in~\cite{Armendariz-Picon:2003knj} with a similar conclusion. Yet another possibility is that the inflationary state is a thermal distribution of particles~\cite{Bhattacharya:2005wn,Berera:1995ie,Cutrona:2026mxm}, such as inflaton quanta and other particles in thermal equilibrium. 
While the predictions for the inflationary observables can significantly change in this case, an initial inflationary period is generically used as a dynamical origin of the cosmological plasma and, if so, precede the generation of temperature.

It is also appropriate to make a further remark.
Here we have constructed a manifestly real classical Hamiltonian for scalars, see Eqs.~\eqref{Hscalar} and~\eqref{Htot}, which can be promoted to a Hermitian Hamiltonian after quantization. 
Terms which mix different variables, like $\Pi_\Psi \Psi$, can be rendered manifestly Hermitian by performing appropriate substitutions, e.g.~$\Pi_\Psi \Psi \to (\Pi_\Psi \Psi+\Pi_\Psi \Psi)/2$, which lead to the same classical limit.
The same is true in the tensor sector: starting from Eq.~\eqref{MomLag} and~\eqref{Pih} one can easily construct a real classical Hamiltonian, which becomes Hermitian after quantization. Although this Hamiltonian depends explicitly on time, the corresponding time-evolution operator $U$ is isometric and, of course, linear, which ensures $U^\dagger U=1$.  Note that isometricity automatically guarantees that quantum-mechanical probabilities are conserved under time translation, which is the necessary physical unitarity condition.

Finally, let us discuss some additional possible outlook. 
\begin{itemize}
\item  When the bound in~\eqref{boundG} and~\eqref{LamEta} are not satisfied it would be interesting to classify the leading higher-derivative operators contributing at $q/a \sim H$ to understand which operators dominate in the slow-roll approximation and explore the implications of these operators for the inflationary observables.  
\item The quantization procedure based on the Dirac formalism employed in this work is computationally involved but highly versatile. 
\begin{itemize}
\item This approach could be extended to higher-order perturbative corrections. For example, one can apply this in the future to cosmological correlators~\cite{Weinberg:2005vy,Armendariz-Picon:2008pwa,Armendariz-Picon:2014xda}. In this context one could also, among other things, estimate the lifetime of excited  states to determine whether they could be viable candidates as inflationary states, without performing the sub-horizon limit.
\item Moreover, this approach could  be extended to unimodular versions of gravity where the spacetime volume is not a dynamical degree of freedom~\cite{Weinberg:1988cp} and therefore extra constraints appear in the quantum theory~\cite{Barvinsky:2019agh,Barvinsky:2019qzx,Salvio:2024vfl}.
\end{itemize}
\end{itemize}

 \vspace{0.7cm}
 
 \subsection*{Acknowledgments}
The authors thank Robert Brandenberger, Alessio Notari and Alfredo Urbano  for useful discussions.

\appendix

\section{List of Poisson brackets and Dirac brackets}\label{List of Dirac brackets}

Here we present the Poisson brackets between the canonical variables and the constraints and  the derivation of the  Dirac brackets between all canonical variables, which are needed to perform the canonical quantization for scalars through the Dirac prescription in~\eqref{eq:DiracP}. Therefore, this appendix provides the commutator between an arbitrary pair of canonical variables.

Using the explicit expressions for the two constraints we have in Eqs.~\eqref{C1ex}-\eqref{eq:C2H}, all equal-time Poisson brackets between the canonical fields and $C_1$ and $C_2$ are given by
\bea
    \left\{\varphi^i(\vec{x}),C_1(\vec{y})\right\} &=& 0, 
    \\ \left\{\varphi^i(\vec{x}),C_2(\vec{y})\right\} &=& -\phi'^i\, \delta^{(3)}(\vec{x}-\vec{y}), \\
    \left\{\Pi_{\varphi i}(\vec{x}),C_1(\vec{y})\right\} &=& -3a^2K_{ji}\phi'^j\, \delta^{(3)}(\vec{x}-\vec{y}),
    \\ \left\{\Pi_{\varphi i}(\vec{x}),C_2(\vec{y})\right\} &=& \left(a^4U_{ ,i} -\frac{a^2}{2}K_{mn,i}\  \phi'^m\phi'^n\right) \delta^{(3)}(\vec{x}-\vec{y}),\\ 
    \left\{\Psi(\vec{x}),C_1(\vec{y})\right\} &=& \delta^{(3)}(\vec{x}-\vec{y}),
    \\ \left\{\Psi(\vec{x}),C_2(\vec{y})\right\} &=& \mathcal{H}\, \delta^{(3)}(\vec{x}-\vec{y}), \\
     \left\{\Pi_\Psi(\vec{x}),C_1(\vec{y})\right\} &=& 0,
    \\ \left\{\Pi_\Psi(\vec{x}),C_2(\vec{y})\right\} &=& 2a^2\bp^2\left[3(\mathcal{H}^2-\mathcal{H'})-\vec\nabla^2_{x}\right] \delta^{(3)}(\vec{x}-\vec{y}).
\eea

With these, together with the definitions of the Dirac brackets in \eqref{eq:DiracP} and the inverse $C$ ``matrix" in \eqref{eq:Cmatrix}, we obtain the following equal-time Dirac brackets: 
\be
    \left\{\varphi^i(\vec{x}),\Pi_{\varphi j}(\vec{y})\right\}_D = \delta^i_j\ \delta^{(3)}(\vec{x}-\vec{y}) +3a^2K_{jl}\phi'^i\phi'^l\ C^{21}(\vec{x},\vec{y}) = \delta^i_j\ \delta^{(3)}(\vec{x}-\vec{y}) -\frac{3}{2\bp^2}K_{jl}\phi'^i\phi'^l\  G(\vec{x},\vec{y}), \nonumber
    \ee
    \be
    \left\{\Psi(\vec{x}),\Pi_\Psi(\vec{y})\right\}_D = \delta^{(3)}(\vec{x}-\vec{y}) + 3(\mathcal{H}^2-\mathcal{H}')G(\vec{x},\vec{y})- \int d^3z\ d^3w\ \delta^{(3)}(\vec{x}-\vec{z})\ G(\vec{z},\vec{w})\vec\nabla^2_w\ \delta^{(3)}(\vec{w}-\vec{y}). \nonumber
\ee
Integrating over $\vec z$ and $\vec w$ and using Property \eqref{eq:LaplacianGreen} of $G(\vec{z},\vec{w})$ and the background Eq.~\eqref{eq:phi'2MF}:
\begin{align} 
    &= \delta^{(3)}(\vec{x}-\vec{y}) -\delta^{(3)}(\vec{x}-\vec{y}) + 3(\mathcal{H}^2-\mathcal{H}')\ G(\vec{x},\vec{y}) =\frac{3K_{ij}}{2\bp^2}\phi'^i\phi'^jG(\vec{x},\vec{y}).
\end{align}
Similar procedures can be applied to derive the remaining  equal-time Dirac brackets: 
\bea
    \left\{\varphi^i(\vec{x}),\Psi(\vec{y})\right\}_D &=& \frac{\phi'^i}{2a^2\bp^2}G(\vec{x},\vec{y}),\\
    \left\{\Pi_{\varphi i}(\vec{x}),\Pi_\Psi(\vec{y})\right\}_D &=& 3a^2K_{ij}\phi'^j\delta^{(3)}(\vec{x}-\vec{y})-9a^2K_{ij}\phi'^j(\mathcal{H}^2-\mathcal{H}')G(\vec{x},\vec{y}),\\
    \left\{\varphi^i(\vec{x}),\Pi_\Psi(\vec{y})\right\}_D &=& 0,\\ 
    \left\{\Psi(\vec{x}),\Pi_{\varphi i}(\vec{y})\right\}_D &=& \frac{1}{2\bp^2}\left(3\mathcal{H}K_{ij}\phi'^j+a^2U_{,i} -\frac{1}{2}K_{mn,i}\ \phi'^m\phi'^n\right)G(\vec{x},\vec{y}), \label{DB13}\\
   \left\{\Pi_{\varphi i}(\vec{x}),\Pi_{\varphi j}(\vec{y})\right\}_D &=&\frac{3a^2}{2\bp^2} G(\vec{x},\vec{y})\Big[(a^2U_{,i}-\frac{1}{2}K_{mn,i}\phi'^m\phi'^n)K_{jl}\phi'^l  \nonumber\\ &&\hspace{.1cm}-(a^2U_{,j}-\frac{1}{2}K_{mn,j}\phi'^m\phi'^n)K_{il}\phi'^l \Big], \label{DB14}\\
    \left\{\varphi^i(\vec{x}),\varphi^j(\vec{y})\right\}_D &=& \left\{\Psi(\vec{x}),\Psi(\vec{y})\right\}_D =  \left\{\Pi_\Psi(\vec{x}),\Pi_\Psi(\vec{y})\right\}_D=0.
\eea

\section{Symplectic form}
\label{appendix:SymplecticForm}
We show here that the antisymmetric bilinear form  in \eqref{eq:Symplecticform} is non-degenerate iff the matrix $\Omega$ is non-degenerate. A generic bilinear form $W$ acting on $\mathscr{V}\times \mathscr{V}$ is said to be non-degenerate when $W(X,Y)=0,\ \forall \ X$, implies $Y=0$. 

First note that, given a basis $\left\{e_1, e_2,\dots, e_{2N}\right\}$ in $\mathscr{V}$, the non-degeneracy condition is equivalent to the non-degeneracy of the matrix $\mathcal{W}$ with elements
\begin{equation}
W_{rs} \equiv W(e_r,e_s).
\end{equation}
Indeed, a general vector $X\in \mathscr{V}$ is written in terms of its components $X_r$ as 
\begin{equation}
X =  \sum_r X_r e_r 
\end{equation}
and so  
\be W(X,Y)= \sum_{rs} X_rW_{rs} Y_s \ee
and if we now assume this is zero $\forall\ X$, we have $W_{rs}Y_s = 0$, which implies $Y=0$ iff the determinant of $\mathcal W$ is non-zero.
 
 Now  the specific matrix $\mathcal W$ with elements $W_{rs}$ defined by Eq.~\eqref{eq:Symplecticform} can be written as a matrix product as follows:
\begin{equation} \mathcal W = 
\begin{pmatrix}
 \tilde\varphi^1_1 & \tilde\varphi^2_{q1} & \cdots & \tilde\varphi^N_{q1} & \tilde\Pi^1_{q1} & \cdots & \tilde\Pi^N_{q1}\\
  \tilde\varphi^1_{q2}& \ddots \\
  \vdots \\
  \tilde\varphi^1_{q2N}
\end{pmatrix}
\ \Omega \ 
\begin{pmatrix}
 \tilde\varphi^1_{q1} & \tilde\varphi^1_{q2} & \cdots & \tilde\varphi^1_{q2N} \\
  \tilde\varphi^2_{q1}& \ddots \\
  \vdots \\
  \tilde\varphi^N_{q1}\\
  \tilde\Pi^1_{q1}\\
  \vdots \\
  \tilde\Pi^N_{q1}
\end{pmatrix}.
\end{equation}
Note that the determinant of the first matrix is non-zero, since a vanishing determinant would imply that one row is a linear combination of the others, but this cannot happen as the rows correspond to {\it independent} solutions of the equations of motion. An analogous argument applies to the last matrix, which is the transpose of the first one. Then, the determinant of $\mathcal W$ is non-zero iff $\Omega$ is non-degenerate.

\vspace{1cm}

 \footnotesize
\begin{multicols}{2}

\end{multicols}

  \end{document}